\documentstyle[prb,aps,twocolumn,rotating,graphics]{revtex}
\begin{document}
\def\luni{LuNi$_2$B$_2$C}
\def\yni{YNi$_2$B$_2$C}
\def\luy{Y$_x$Lu$_{1-x}$Ni$_2$B$_2$C}
\def\lunine{Y$_{0.1}$Lu$_{0.9}$Ni$_2$B$_2$C}
\def\lufive{Y$_{0.5}$Lu$_{0.5}$Ni$_2$B$_2$C}
\def\lutwo{Y$_{0.8}$Lu$_{0.2}$Ni$_2$B$_2$C}
\def\lan{La$_3$Ni$_2$B$_2$N$_{3-\delta}$}
\def\rni{$R$Ni$_2$B$_2$C}
\def\[{\begin{eqnarray}}
\def\]{\end{eqnarray}}
\def\<{\langle}
\def\>{\rangle}

\title{Superconducting properties of \luy\ and 
\lan: \\A comparison between experiment and Eliashberg theory}
\author{S. Manalo, H. Michor, M. El-Hagary, and G. Hilscher}
\address{Institut f\"ur Experimentalphysik, Technische Universit\"at Wien,
Wiedner Hauptstra\ss e 8-10, A-1040 Wien, Austria}
\author{E. Schachinger}
\address{Institut f\"ur Theoretische Physik, Technische Universit\"at Graz, 
Petersgasse 16, A-8010 Graz, Austria}
\date{\today}
\wideabs{
\maketitle
\begin{abstract}
Specific heat and resistivity measurements were performed on polycrystalline 
samples of the solid-solution  
\luy\ in order to determine thermodynamic properties such as the 
specific-heat difference $\Delta C$, the thermodynamic critical field 
$H_c(T)$, as well as the upper 
critical field $H_{c2}(T)$. 
These properties were analyzed within the Eliashberg theory including 
anisotropy effects, 
yielding electron-phonon coupling anisotropy parameters $\< a_{\bf 
k}^2\>$ ranging between $0.02$ and $0.03$ for the whole series, and  
Fermi velocity anisotropy parameters of $\< b_{\bf k}^2\> = 0.245$--$0.3$.
Excellent agreement between theory and experiment was achieved for 
these parameters, the Sommerfeld constant $\gamma$ and model 
phonon spectra determined from specific heat measurements. 
An analysis of the previously investigated boronitride \lan\ for comparison 
revealed the electron-phonon anisotropy to be of great significance in 
describing its thermodynamic properties and the calculations 
yielded $\< a_{\bf k}^2 \> \simeq 0.08$ and $\< b_{\bf k}^2 \> \simeq 
0.245$. The $T_c$-behavior within the series \luy\ is discussed in terms of 
%coupling and impurity effects, and 
he density of states at the Fermi level $N(0)$.  
\end{abstract}
}

\section{INTRODUCTION}

The transition-metal borocarbide superconductors $R$Ni$_2$B$_2$C with 
transition temperatures comparable to 
those of the A-15 compounds (e.g.~$R$ = Lu  with $T_c \simeq 16.5$\,K, 
Nb$_3$Ge with $T_c \simeq 23$\,K) are a subject of broad interest 
for research on intermetallic superconductors. Siegrist {\it et 
al.\/}\cite{sieg} reported the crystal 
structure of the $R$Ni$_2$B$_2$C superconductors to be a filled version 
of the ThCr$_2$Si$_2$-type structure stabilized by the incorporation of 
carbon, where Ni$_2$B$_2$ layers are separated by $R$C layers. In 
the related compound  \lan,  three LaN-planes separate 
the Ni$_2$B$_2$ layers. Despite of the layered structure  
reminiscent of the cuprate superconductors, the electronic 
structure of the single $R$C-layer borocarbides as well as the 
triple LaN-layer boronitride is three-dimensional.\cite{matth}$^,$\cite{sing} 
Nickel-site substitutions on both the 
borocarbides 
and the boronitride revealed similar electronic properties of 
the bands related to the 3$d$-electrons.\cite{grml} 
 
At a first glance, the thermodynamic properties of \lan\ seem to be close 
to the weak coupling BCS-predictions,\cite{lanpap} but standard single-band 
BCS-theory cannot explain the pronounced upward curvature of the upper critical 
field $H_{c2}(T)$ close to $T_c$.
If applied to the borocarbides, it fails to describe both the thermodynamic 
properties and the upper critical field
obtained from experiment. Shulga {\it et al.\/}\cite{shulga} analyzed the 
upper critical field $H_{c2}(T)$ of \luni\ and \yni\ in terms of
the Eliashberg-theory using an isotropic two-band model. An
isotropic
single band model cannot reproduce the positive curvature near 
$T_c$
apparently, because of the dominant role of anisotropy effects in this
system. Recently, Dugdale {\it et al.\/}\cite{dugdale} presented an 
experimental investigation on the Fermi surface of \luni, revealing that 
it consists of three sheets. 
  
Freudenberger {\it et al.\/}\cite{freude} studied the solid solution 
\luy\ and showed, that $T_c$ exhibits a minimum at about $x 
\simeq 0.5$ with a $T_c$-depression of about $1$\,K.
They pointed out, that this feature cannot be described alone by 
disorder effects with the residual resistivity ratio as a measure, and
tentatively supposed the electron phonon coupling strength $\lambda$ to be 
the origin of this behavior.
 
We present in this paper investigations of the thermodynamic 
properties and of the upper critical field $H_{c2}(T)$ of the series \luy\ 
and the boronitride \lan. The critical temperature, the 
specific heat, and the temperature dependence of the upper critical 
magnetic field were measured for all samples of the series \luy\ and the 
results, including previous measurements of \lan\ 
(Ref.~\onlinecite{lanpap}), were 
analyzed using an anisotropic model of the $s$-wave Eliashberg-formalism.
Considering the change in mass from Y to Lu in the series and different 
lattice properties of both systems, we 
used model phonon spectra calculated from the phonon contribution to 
the specific heat and found these to be sufficient in 
describing both the thermodynamic properties and the upper critical field of 
\lan\ and \luy. 
An analysis of the coupling strength and electronic density of states in 
\luy\ gives an insight on the origin of the dip in $T_c(x)$. 
Sec.\,II presents details of sample 
preparation and measuring techniques and Sec.\,III shows the theoretical 
background used for the analysis. In Sec.\,IV, an analysis of the 
experimental data in terms of the
anisotropic Eliashberg theory is discussed and possibilities for the 
$T_c$-reduction are presented. Conclusions are drawn in Sec.\,V.  

\section{ EXPERIMENTAL} 

Polycrystalline samples of \luy\ were synthesized
on a water-cooled copper grove by high frequency induction melting under
argon atmosphere. The starting materials are rare earth ingots ($R$ = Y,
Lu) (Strem chemicals, USA: 99.9\%), Ni ingots (Strem chemicals, USA:
99.9\%), crystalline boron (Starck, Germany: 99.5\%) and carbon ingots
(Starck, Germany: 99.99\%). A good homogeneity was obtained by performing
the following three steps of sample preparation: (i) Nickel and boron were
melted two times. (ii) Rare-earth and carbon ingots were alloyed together
while compensating the carbon loss (1-2\%) during the remelting stage.
(iii) The two precursor alloys NiB and $R$C were melted together. The
buttons were broken and remelted for at least 12 times, and finally
annealed at $1020^\circ$C for at least one week in evacuated quartz
tubes.  The phase purity of the samples was checked at room temperature
by applying X-ray diffraction (XRD) in a Guinier-Huber camera using
germanium as an internal standard. The X-ray photographs of the samples,
indexed on the basis of the tetragonal crystal structure of \luni\ (space
group $I4/mmm$), show that the samples of the solid solution \luy\ are
almost single phase with traces of Y$_{1-x}$Lu$_x$B$_2$C$_2$ and
Y$_{1-x}$Lu$_x$Ni$_4$B.  According to the x-ray line intensities and
optical micrograph investigations the total amount of these secondary
phases is about 2--5$\%$ in all samples investigated.

Specific heat
measurements in the temperature range 1.5\,K-160\,K and magnetic fields up
to 9\,T were carried out on 1-2\,g samples employing a quasi-adiabatic
step-heating technique. The sample holder consists of a thin sapphire disc
($m\sim 0.2$\,g) with a strain gauge heater and a CERNOX temperature
sensor. The field calibration of the latter has been performed against two
GaAlAs resistivity thermometers and a capacitive SrTiO$_3$ sensor.  Four
point resistivity measurements were performed in fields up to 9\,T in
order to determine $H_{c2}(T)$ by means of a midpoint criterion.  The
transition widths according to a 10$\%$ and 90$\%$ criterion are indicated 
by error bars in the figures containing the upper critical field obtained 
from the experiment.

%%%new

The characterization of the polycrystalline \luy\ samples with respect to their
residual resistance $\rho_0$ and the room temperature residual resistance ratio (RRR) 
turned out to be rather problematic, because our well annealed samples are rather brittle 
and partly textured. Thus, we had difficulties to obtain reproducible and reliable values 
of the RRR ranging between 7 and 41. We obtained a very high RRR value of 41 for \luni\ 
(twice as large as typical single crystal values, see e.g.\ Ref.~\onlinecite{rathna})
but a much lower value of just 12
for \yni\ although its thermodynamic mean transition temperature $T_c=15.56$\,K 
(see below) is even slightly higher than $T_c=15.4$\,K of a high quality \yni\ single 
crystal\cite{nohara} with RRR$=37.$ The upper critical fields $H_{c2}(T)$ match to
each other despite of the different RRR values (compare Fig.~11 of this work with Fig.~4
of Ref.~\onlinecite{nohara}).
In the latter paper Nohara {\it et al.}\cite{nohara} reported that the Sommerfeld 
value $\gamma$ of clean limit \yni\ and \luni\ exhibits a $\sqrt{H}$ dependence which
changes to a linear relation $\gamma\propto H$ in the dirty limit. We note,
that we found an approximate $\gamma\propto\sqrt{H}$ dependence in all our samples including
\lan, although the RRR shown in Ref.~\onlinecite{lanpap} is only 3, similar to
the "dirty" Y(Ni$_{0.8}$Pt$_{0.2}$)$_2$B$_2$C for which $\gamma\propto H$ was reported.
It is obvious that the grain boundaries of the polycrystalline samples strongly 
reduce the RRR value especially in \lan. Further arguments for the extrinsic
origin of the large residual resistance $\rho_0$ of \lan\ will be given in context 
with analysis of the experimental data in terms of the anisotropic Eliashberg theory 
(see Sec.~VI\,E).  
    
%%%

\section{Theoretical background} 

Anisotropy effects have to be considered in the description of the 
experimental results for $H_{c2}(T)$ in order to obtain a satisfactory 
agreement between theory and experiment. 
Daams and Carbotte\cite{daamcar} applied the separable model introduced by 
Markovitz and Kadanoff\,\cite{marka} to describe an anisotropic 
electron-phonon interaction spectral function within Eliashberg theory:
\begin{equation}  
[\alpha^2F(\omega)]_{\bf k, k'} = (1+a_{\bf k})\alpha^2F(\omega)(1+a_{\bf k'})\,, 
\label{eq1} 
\end{equation}
where {\bf k} and {\bf k'} are the incoming and outgoing quasi-particle
momentum vectors in the electron-phonon scattering process and $a_{\bf k}$
is an anisotropy function describing the deviation of the anisotropic
spectral function,
$[\alpha^2F(\omega)]_{\bf k, k'}$, from the isotropic
one, $\alpha^2F(\omega)$, in the direction of {\bf k}. It has the 
important feature that its Fermi surface average $\<a_{\bf k}\> = 0$ and 
as anisotropy effects are rather small it is sufficient to keep only the 
mean-square anisotropy $\<a_{\bf k}^2\>$ as the important 
anisotropy parameter. 
 
The theory of $H_{c2}(T)$ for anisotropic polycrystalline superconductors
in a separable model scheme was developed by Prohammer and 
Schachinger.\cite{pro} It employs the separable ansatz for the 
anisotropy of the electron-phonon interaction and the ansatz
\begin{equation}
v_{F,{\bf k}} = (1 + b_{\bf k})\< v_F \>\,,
\end{equation}
which describes the anisotropy of the Fermi velocity; $b_{\bf k}$ is 
an anisotropy function defined the same way as $a_{\bf k}$.
The upper critical field is then described by the set of
equations (27-30 in Ref.~\onlinecite{weber}) and thermodynamic 
properties of the borocarbides were calculated using Eqs. (33-37) 
of Ref.~\onlinecite{weber}.  
$N$-band models have been extensively studied as a tool to describe
anisotropic features of superconductors. Shulga {\it et al.\/}\cite{shulga} 
applied the two-band model $H_{c2}$
equations\cite{pro} to their experimental results and were able to 
describe the experimental data with the corresponding equations. 
Thus, the separable model employed in this work can be described in its 
simplest form by a Fermi surface split into two half-spheres of equal weight
\begin{equation}
P(a)=\delta(-a)/2 + \delta(a)/2\,,
\label{disfunc}
\end{equation}
with radii $r\pm a$, if $r$ is the radius of the equivalent isotropic
Fermi sphere.\cite{niel} Using the Fermi surface harmonics (FSH) notation
introduced by Allen,\cite{allen} Daams\cite{daams} 
observed that this
separable model was equivalently described by a restriction to 
zeroth-order FSH in each of the two subregions of the Fermi 
surface. According to 
her work, the separable model applied in our analysis 
corresponds to a two-band model in which the two Fermi surface regions 
have equal weight. It is of course also possible to define different 
weights for the two regions, thus changing the distribution function 
(\ref{disfunc}), but Daams\cite{daams} demonstrated that in the case of
weak anisotropy the influence of different weights in a separable
model is of negligible significance for the thermodynamics of
anisotropic superconductors.
 
In the case of the upper critical field any deviation from the 
equal weight configuration causes $H_{c2}(T)$ 
to approach the isotropic case as is demonstrated in Fig.~\ref{luisoan}.
The deviation function in the insert of Fig.~\ref{luisoan}
\begin{equation}
D_{H_{c2}}(T/T_c) = D_{H_{c2}}(t) = {H_{c2,a}(t)\over{H_{c2,i}(t)}} - 1
\end{equation}
demonstrates the deviation of the upper critical field of an anisotropic 
system $H_{c2,a}(T)$ from the upper critical field of an isotropic, 
equivalent system $H_{c2,i}(T)$. The numerical result for equal weights 
(1:1) was fitted to the upper critical field of 
\luni\ to fix the parameters $\< a_{\bf k}^2 \>,\ \< b_{\bf k}^2\>,\ {\rm 
and}\ \< v_F\>$. The mean Fermi velocity $\< v_F\>$ and its anisotropy 
parameter $\< b_{\bf k}^2\>$ were used to fit the experimental data near 
$T_c$, and $\< a_{\bf k}^2 \>$ was changed to describe $H_{c2}(T)$ at 
lower temperatures. With these parameters, we calculated the upper critical 
field for different weights of the Fermi sheets (1:$n$) to investigate the 
change in the behavior of $H_{c2}$. Obviously, $H_{c2,a}$ approaches  
$H_{c2,i}$ as one of the two sheets becomes dominant in weight and this 
is indicated by a flattening of $D_{H_{c2}}(t)$. 
 
The relative signs of $a_{\bf k}$ and $b_{\bf k}$ in the same 
Fermi-surface sheet is also of importance to the analysis of $H_{c2}$.
With same signs of $a_{\bf k}$ and $b_{\bf k}$ in same sheets, 
$H_{c2,a}(T)$ is reduced compared to $H_{c2,i}(T)$ (Fig.~\ref{luiso}). 
Opposite signs 
within same sheets give rise to an enhancement of $H_{c2,a}$ 
over $H_{c2,i}$. In this case, the reduction of $H_{c2,a}(T)$ due 
to the influence of $\< b_{\bf k}^2\>$ is compensated by 
the electron-phonon interaction anisotropy. 
Bandstructure calculations can give a good estimate for the weight 
and the relative signs of $a_{\bf k}$ and $b_{\bf k}$ in the 
corresponding regions of the Fermi surface. Dugdale {\it et 
al.\/}\cite{dugdale} were able to show that the 
Fermi surface of \luni\ consists of three sheets without giving 
the weights belonging to the Fermi surface regions. The third one is a 
small electron pocket centered at $\Gamma$, which compared to the other two 
sheets can be regarded to have a negligible effect on $H_{c2}$ if 
considered in the theoretical calculations. 
 
With the knowledge of the Fermi surface consisting of two dominant 
sheets and the shape of the experimental upper critical field favouring 
opposite signs of $a_{\bf k}$ and $b_{\bf k}$ within the same sheets, we 
used equal weights for the two Fermi surface regions and 
opposite signs of $a_{\bf k}$ and $b_{\bf k}$ in the following analysis. 
Numerical results for $H_{c2}(T)$ showed that 
in the case of both anisotropy functions having the same signs, 
the mean Fermi velocity $\< v_F\>$ needed to fit the experimental 
data deviates significantly from the value 
determined from the plasma frequency\cite{bommel}. The same effect on 
$\< v_F \>$ is observed with different weights of the Fermi surface regions.
 
Finally, impurities are treated in Born's limit\cite{ambeg} which
assumes the impurities to be randomly distributed and to be of
dilute concentration. In such a limit impurities are characterized
by a scattering rate $t^+$ which is proportional to the
impurities' concentration. Their main effect is the smearing
out of the electron-phonon interaction anisotropy resulting
in a slight reduction of $T_c$,\cite{marka} an enhancement of
$H_{c2}$ at low temperatures, and a reduction of the high
temperature upward curvature of $H_{c2}$ as was demonstrated
by Weber {\it et al.\/}\cite{weber} for Nb.

\section{RESULTS \& DISCUSSION} 

\subsection{Results of the specific heat measurements}

The low-temperature specific heat of \luy\ with $x=0,\ 
0.1,\ $0.5$,\ 0.8$, and $1$ is presented in Figs. 
\ref{ylunull}-\ref{yluphon}. An external field of $9$\,T was 
applied to suppress superconductivity in order to determine the normal state 
heat 
capacity $C_n=C_e+C_{ph} = \gamma T + \beta T^3$, 
where $\gamma$ is the 
Sommerfeld parameter and $\beta$ is related 
to the low-temperature value 
of the Debye temperature. From the specific heat data in the normal and 
superconducting state (Figs.~\ref{ylunull} and 
\ref{ylucel}) we obtained the specific heat difference $\Delta C = C_s-C_n$ 
and used it to calculate further 
quantities as the thermodynamic critical 
field $H_c(T)$ and its deviation function $D(T)$ (Eqs. 
(\ref{hcrit}-\ref{devi})), which are compared to numerical 
results obtained with Eliashberg theory as discussed below. The 
extrapolation of the  
normal state specific heat to $T\rightarrow 0$ 
shown in Fig.~\ref{ylucel} with the range of extrapolation starting at $\sim 30$\,K$^2$
revealed the Sommerfeld constants of the series 
\luy\ ranging from $\gamma = 18.5(2)$\,mJ/mol\,K$^2$ for $x=0.8$ to 
$20.6(2)$\,mJ/mol\,K$^2$ for $x=0$. The values for the other 
compounds within the series are 
$19.7(3)$\,mJ/mol\,K$^2$ ($x=0.1$), 
$18.7(3)$\,mJ/mol\,K$^2$ ($x=0.5$), and $19.7(2)$\,mJ/mol\,K$^2$ ($x=1$). 
A comparison of the Sommerfeld constants of the border phases ($x = 
0,1$) with those of Ref.~\onlinecite{mich} reveals about $\sim 
1$\,mJ/mol\,K$^2$ higher values of $\gamma$ in this work, with about 
the same $T_c$ for \luni\ but a higher one for \yni\ compared to previous 
results. The difference in these properties may be explained by the 
homogeneity range achieved by different methods 
used to prepare polycrystalline materials.\cite{schmidt}  

\subsection{Spectral function and coupling strength}

The large mass difference between Y and Lu causes a significant change of the
low energy phonon frequencies in Y$_{x}$Lu$_{1-x}$Ni$_2$B$_2$C while high 
energy phonon modes are hardly affected by the rare earth substitution.
The latter was shown by inelastic neutron spectroscopy (INS) 
results on Y- and LuNi$_2$B$_2$C reported by Gompf {\it et 
al.\/}\cite{gompf} Similar conclusions were also obtained from the 
lattice heat capacity of $R$Ni$_2$B$_2$C ($R=$ La, Lu and Y) analyzed 
by means of a model 
phonon density of states (PDOS) in a previous paper.\cite{mich} 
The low energy part of our model spectra were well confirmed 
by INS while the high energy part is, of course, 
just a crude description
which does not give the real details. Nevertheless, the moments of these
spectra are in remarkable agreement with those calculated from the
INS results. As the following discussion 
of superconducting properties in terms of  
the Eliashberg formalism requires primarily details of the low energy PDOS
we evaluated the variation of the low energy phonon modes within the series
Y$_{x}$Lu$_{1-x}$Ni$_2$B$_2$C by a similar procedure as explained in 
Ref.~\onlinecite{mich},
i.e.\ by fitting the normal state specific heat measured from 2 to 160\,K
with a model PDOS, $F(\omega),$ which is of the form 
%\widetext
\begin{eqnarray}   
F(\omega)  =  3R \left(\frac{3\omega^2}{\Omega_{\rm D}^3}
\,\theta (\omega)\theta (\Omega_{\rm D}-\omega)\right) +\cr 
\sum_{i=0}^3 g_iR\Bigl(\frac{1}{\sqrt{2\pi\sigma_i^2}}
\exp\left[ -\frac{(\omega -\Omega_{Ei})^2}{2\sigma_i^2}\right]\times\cr 
\theta (\omega -\omega_0)\theta (\Omega_{E3}-\omega)\Bigr). 
\label{f_omega}
\end{eqnarray}
%\narrowtext
In this formula $\theta (x)$ is the well-known step function; 
$\omega_0=4.3$\,meV 
in order to avoid a finite spectral density at zero energy; 
$\sigma_i$ are the widths of the Gaussians and $g_i$ are the weights of these
contributions (i.e.\ the number of the contributing phonon branches).   
The Debye spectrum ($\propto\omega^2$) represents the spectral 
weight 
of the three acoustic branches and the Gaussian contributions account for 
the 
15 optical branches. The free parameters are the Debye cut-off 
$\Omega_{\rm D}$ and the Gaussian peak positions $\Omega_{E_i},$ 
respectively.
In order to fix a high energy limit of the model spectrum, the Gaussian 
contribution with the highest 
energy ($\Omega_{E_3}$) was cut off at the peak position and $g_3$ is 
accordingly 
doubled in equation (\ref{f_omega}). The parameters applied in 
Eq. (\ref{f_omega}) are
summarized in Table~\ref{herwtab}. The only difference with respect to the 
model spectra shown in Ref.~\onlinecite{mich} is the splitting up of 
$\Omega_{E_1}$ with $g_1=1.5$ into
two parts $\Omega_{E_0}$ and $\Omega_{E_1}$ with $g_0=0.3$ and $g_1=1.2,$ 
respectively,
which was motivated by the shape of $F(\omega )$ obtained by INS on 
YNi$_2$B$_2$C 
(see Ref.~\onlinecite{gompf}). On the other hand we now fixed 
$\Omega_{E_3}$ to $103.4$\,meV (equivalent to $1300$\,K$\times k_B$) in 
order to keep the number of parameters small. Only for 
Y$_{0.5}$Lu$_{0.5}$Ni$_2$B$_2$C we
applied a slightly modified spectrum with $g_0=0.7$ and 
$g_1=0.8$ compared
to 0.3 and 1.2 given in Table~\ref{herwtab}, because of the fifty to fifty 
ratio of Y and Lu in this sample. 
     
Adjusting the model spectra to the lattice heat capacity via   
\begin{equation}            \label{einint}
C_{ph}(T)=R\int_{0}^{\infty} {\rm d}\omega\,F(\omega)\frac{\left( 
\frac{\omega}{2T} \right) ^2}{\sinh^2\left( \frac{\omega}{2T} \right) }\,, 
\end{equation}
we obtained the model spectra shown in Fig.~\ref{spek}.
This procedure provides a satisfactory description of the $C_P(T)$  
data from 2\,K up to 160\,K (up to 300\,K for the border phases). 
 
The phonon contribution to the specific heat yields information on 
$F(\omega)$ rather than on $\alpha^2F(\omega)$. Therefore, one is led to 
introduce an assumption on the form of the electron-phonon coupling function 
$\alpha^2(\omega)$. 
Junod {\it et al.\/}\cite{junod} suggested a function $\alpha^2(\omega)\sim 
\omega^s$ in a first approximation and found the exponent to be $s = - 1/2$ 
for a large number of A15 compounds by comparing the 
experimentally determined moments of $\alpha^2F(\omega)$ with tunneling 
data. We calculated the thermodynamic 
properties of the borocarbides using $s=0$, $-1/2$, and $-1$, and found 
the exponent $s = -1/2$ to be the best choice for the theoretical 
analysis as in the case of the A15 compounds. It was impossible 
to describe the experimental data using $s=0$ because the theoretical 
deviation function $D(t)$ already lies below the the experiment for 
an isotropic system, as a consequence of the large gaussian contribution
above the low temperature Debye frequency $\Omega_D^{LT}$, 
and is expected to get more negative for an anisotropic system. 
In the case of $s = -1$ the calculated deviation function is significantly 
larger than the experimental one and a consistent description of the 
thermodynamic properties and $H_{c2}$ is impossible because of the large 
anisotropy parameter $\<a_{\bf k}^2\>$ necessary to describe the 
thermodynamic properties. 
 
The spectra were cut off at the local minimum of the model phonon spectra 
at $\omega\sim 68$\,meV. Starting at this energy, 
the phonon spectra determined by Gompf {\it et al.}\cite{gompf} using 
inelastic neutron-scattering techniques have a wide 
interval, where the PDOS is zero. 
The next optical 
contributions to $F(\omega)$ start at $\sim 100$\,meV, and it is not very 
likely that electrons still considerably couple to phonons at these 
energies. 
 
With the coupling function having the form $\alpha^2(\omega) = 
a\,\omega^{-1/2}$, the model spectra shown in Fig.~\ref{spek} were 
weighed with a function $\omega^{-1/2}$ and were rescaled with a constant 
$a$ to give the corresponding critical temperatures $T_c$ of the samples 
for a fixed pseudopotential of $\mu^* = 0.13$\,. 
The model spectral functions give electron-phonon interaction mass 
enhancement factors 
\begin{equation}
\lambda 
= 2\int_0^\infty {\rm d}\omega\,{\alpha^2F(\omega)\over{\omega}} 
\label{lambda}
\end{equation}
between 1.02 ($x=1$) 
and 1.22 ($x=0$). 
$\lambda$ does not behave as the critical temperature and 
descends monotonically from \luni\ to\ \yni\ as a function of $x$, without 
showing a minimum, which indicates that it cannot account alone, if at all,
for the depression of $T_c$ within the series. 

\subsection{Thermodynamic properties and $H_{c2}(T)$}

With the set of parameters given in Table~\ref{table2}, the thermodynamic
properties were calculated for all samples. The number of atoms
in a volume of 1\,cm$^3$, $n_A$, was calculated from lattice parameters 
$a$ and $c$ of the tetragonal structure. 
The Sommerfeld constants obtained from the experiment, $\gamma_{exp}$,   
are slightly greater than the theoretical values, $\gamma_c$, needed to fit 
the experimental 
results for the thermodynamic critical field $H_c(T)$. The latter is 
calculated from the experimental data by integrating the difference 
between the zero-field and $9$\,T measurement, i.e.~any normal state 
contribution to the zero-field heat capacity cancels out in $\Delta C = 
C_s - C_n\ \hat=\ 
C(0\,{\rm T}) - C(9\,{\rm T})$ which is used in the analysis below. 
Nevertheless, a normal state contribution, which is eliminated 
in $\Delta C$, will strongly influence the experimental result for 
the temperature dependence of the 
electronic specific heat in the superconducting state $C_{eS}(T)$, 
obtained by subtracting 
the phonon contribution from the zero-field data. We note, that the 
$\gamma$-value due to a small amount of impurity phase obtained by 
extrapolating the zero field data in a $C/T$ {\it vs.\/} $T^2$ plot to 
$T\rightarrow 0$ yields e.g. for \yni\ $\gamma_{imp} \simeq 
0.4(2)$\,mJ/mol\,K$^2$ which 
is significantly smaller than $\gamma_{exp}-\gamma_c \sim 
1.1$\,mJ/mol\,K$^2$. 
For all other samples, $\gamma_{imp}$ is found to be even closer to zero, 
but $\gamma_{exp}-\gamma_c$ is still about 1\,mJ/mol\,K$^2$ 
(see Fig.~\ref{gamma}). Thus, the average difference of about $5 \%$ may be 
attributed to an intrinsic effect, namely to the contribution of the small 
electron pocket, 
for which de Haas-van Alphen measurements performed by Terashima {\it et 
al.\/}\cite{terashima} revealed the superconducting gap to be much 
smaller than on other parts of the Fermi surface. For a further 
discussion of the temperature dependence of $C_{eS}(T)$ see below (Sec. 
IV F). 
 
In the description of the thermodynamics and the upper critical field 
of \luy\ the model spectra of Fig.~\ref{spek} were used in the numerical
solution of the Eliashberg equations in the clean limit (impurity
scattering rate $t^+ = 0$). This is of course only a rough approximation
and the corresponding scattering rate $t^+$ should be determined from a
$T_c$ reduction with increasing concentrations of Born limit momentum
scattering impurities.\cite{weber} As it was not possible to perform this
type of experiment with the present samples we investigated the influence
of impurity concentration theoretically and realized that in order to stay
within the constraints of the known experimental data $t^+$ had
to be rather small and therefore we regarded the clean limit to
be an acceptable approximation. As the coherence length is rather
short in these materials the clean limit is always a good approximation. 
The results of our theoretical study of different spectral functions
$\alpha^2F(\omega)$ and of the influence of impurities on the
thermodynamics as well as on $H_{c2}$ will be the topic of a
separate, forthcoming publication.
  
Fig.~\ref{luall} depicts the comparison between experimental results and 
theoretical predictions for \luni. The top frame shows 
the specific heat difference between superconducting and normal state
$\Delta C(T) = C_s(T)-C_n(T)$. Theoretical calculations were done for an
isotropic ($\<a_{\bf k}^2\> = 0$) and an anisotropic case ($\<a_{\bf 
k}^2\> = 0.03$), and results with anisotropy parameters between these 
values lie in between the numerical results of Fig.~\ref{luall}(a-c). 
Considering impurity scattering would bring the results of an 
anisotropic calculation closer to the isotropic case.  
 
The thermodynamic critical field $H_c(T)$ is calculated from the free
energy difference $\Delta F$
\begin{equation}
H_c(T) = \sqrt{2\Delta F/\mu_0}\,,
\end{equation}
and the specific heat difference $\Delta C$ is related to $\Delta F$ 
through \begin{equation}
\Delta C(T)= -T{\partial^2\Delta F(T)\over{\partial T^2}}\,.
\label{hcrit}
\end{equation} 
Strong coupling and anisotropy effects can be observed in the 
deviation function of the thermodynamic critical field
\begin{equation}
D(T) = {H_c(T)\over{H_c(0)}} -
\Bigl[1-\Bigl({T\over{T_c}}\Bigr)^2\Bigr]\,.
\label{devi}
\end{equation}
Error bars of the experimental deviation function result 
from experimental error bars of the specific heat data and 
primarily from the extrapolation of $H_c(T)$ to $T\rightarrow 0$. 
Anisotropy effects push the deviation function towards the weak
coupling regime as shown in Fig. \ref{luall}(c). Even with the large 
error bar attached to the experimental data, it becomes obvious that the 
BCS result is far off with its minimum at about $-3.7\%$.
 
The upper critical field $H_{c2}(T)$ was fit to the experimental data by
describing the positive curvature at $T_c$ with the Fermi velocity
anisotropy parameter $\<b_{\bf k}^2\>$ at a fixed mean Fermi velocity
$\<v_F\>$, and by changing the anisotropy parameter $\<a_{\bf k}^2\>$ of the
electron phonon coupling strength to describe the behavior of the
experimental curve at lower temperatures. The value of $\<v_F\>$ was taken
from the experimental result for the plasma frequency $\hbar\omega_{pl} =
\sqrt{4\pi e^2v_F^2N(0)/3} = 4.0$\,eV (see Ref.~\onlinecite{bommel}) which 
results in a mean
Fermi velocity $\<v_F\>$ of $\simeq 0.28\times 10^6$~m/s for 
\luni.\cite{shulga} 
A reasonable fit of the experimental data with this value is achieved 
with $\<a_{\bf k}^2\> = 0.02$ and $\<b_{\bf k}^2\> = 0.25$. According 
to the thermodynamic properties, it would as well be possible to apply 
a higher anisotropy parameter, i.e. $\<a_{\bf k}^2\> = 0.03$. Error bars 
of the plasma frequency were not given in Ref.~\onlinecite{bommel}, and a 
change of $0.01\times 10^6$~m/s in $\<v_F\>$ is likely to be within 
the experimental error range. For an anisotropy parameter of $\<a_{\bf 
k}^2\> = 0.03$, one has to reduce $\<b_{\bf k}^2\>$ and enhance
$\<v_F\>$ to compensate the change as depicted in the bottom frame 
of Fig.~\ref{luall}. The error bars of the experimental upper critical field  
result from the width of the transition in the resistivity data. 
 
Figs.~\ref{cps}, \ref{hcs}, \ref{hc2s}, and \ref{devis} show the results
for the samples with $x = 0.1,\ 0.5,\ 0.8$, and $1$. Thermodynamic 
properties
and $H_{c2}$ of the whole series
can be described with anisotropy parameters $\<a_{\bf k}^2\>$ between
$0.02$
and $0.03$, which is relatively small compared to the Fermi velocity
anisotropy $\<b_{\bf k}^2\>$ ranging from $0.245$ to $0.3$ and is
comparable
to the electron-phonon anisotropy parameters of LaAl$_2$ 
(Ref.~\onlinecite{vlcek}, $\<a_{\bf k}^2\> = 0.01$), Nb  
(Ref.~\onlinecite{weber}, $\<a_{\bf k}^2\> = 0.0335$) and is smaller than 
the anisotropy parameter of Nb$_3$Sn (Ref.~\onlinecite{schach}, 
$\<a_{\bf k}^2\> = 0.08$). The Fermi 
velocity anisotropy parameters of \luy\ are obviously higher than 
those of LaAl$_2$ ($\<b_{\bf k}^2\> = 0.16$), Nb ($\<b_{\bf k}^2\> = 
0.118$), and Nb$_3$Sn ($\<b_{\bf k}^2\> = 0.13$). In comparing the upper
critical field of \yni\ to that of \lufive\ it is obvious that the
positive curvature of $H_{c2}$ at $T_c$ is rather pronounced in both
samples. If the $T_c$ reduction had been caused by impurity scattering the
positive curvature of $H_{c2}$ at $T_c$ would have been gradually reduced
with increasing impurity concentration as was
demonstrated by Weber {\it et al.\/}\cite{weber} in their analysis of
polycrystalline Nb. Such an effect can certainly not be observed
in our case.

\subsection{$T_c$-reduction in the series \luy}

The reduction of $T_c$ within the series cannot be described with 
impurity scattering in terms of Eliashberg 
theory because this would require randomly distributed scattering centers. A 
substitution of Lu with Y or vice versa produces inhomogeneities of the 
periodic potential only on the rare earth lattice 
site, so that a random distribution can only be expected with  
small substitutions up to $x = 0.1$\,.  If the   
$T_c$-reduction should be explained by inelastic impurity scattering  
expressed by the scattering rate $t^+$, it 
is restricted to the condition that the electronic properties of the 
system, as e.g.\ $N(0)$, remain unchanged, which is not the case in 
\luy\ (See Fig.~\ref{tcargh}). Thus, the $T_c$-reduction within the series 
cannot be attributed to inelastic impurity scattering expressed by 
$t^+$.

The density of states $N(0)$ (Fig.~\ref{tcargh}) was calculated using the 
bare Sommerfeld constant $\gamma^*=\gamma/(1+\lambda)$ and the number of atoms per
cm$^3,$ $n_A,$ in the approximation $N(0) \propto 
n_A\gamma^*$. Unlike the electron-phonon coupling factor $\lambda$, 
$\gamma$ behaves like $T_c$ as a function of $x$, therefore accounting 
for a similar behavior of $N(0)$. 
Hence, as $\lambda$ decreases with 
decreasing Lu-content, $N(0)$ shows a minimum between $x=0$ and $1$. 
Obviously, the behavior of $T_c$ results from an 
interplay of the coupling strength of electrons to phonons 
and the density of states at the Fermi level. The latter is most likely 
due to a band broadening arising from the $R$-ion size mismatch which 
washes out the maximum of the electronic density of states at the Fermi energy.
The $N(0)$ values of the unalloyed compounds \yni\ (4.14 states/eV\,f.u.) 
and \luni\ (3.94 states/eV\,f.u.) agree closely with the values of 
4.31 and 4.06 states/eV\,f.u., respectively, reported from density functional
calculations by Divis {\it et al.}\cite{divis}  

\subsection{Thermodynamics and $H_{c2}$ of \lan}

%%% new
The model spectrum of \lan\ (Fig.~\ref{lana2f}) was obtained in a similar 
manner as the spectra of the borocarbides (Ref.~\onlinecite{mich}) and 
is cut off at the local minimum at $\sim 50$\,meV. After applying a 
coupling function $\alpha^2(\omega) \sim \omega^s$ with $s = -1/2$, the 
spectrum was rescaled with a constant factor to give the critical temperature 
$T_c=11.73$\,K and a fixed pseudopotential $\mu^* = 0.13$. 
This results in a coupling strength of $\lambda = 1.02$\,. 
With the Sommerfeld constant $\gamma = 24$\,mJ/mol\,K$^2$, the 
thermodynamic properties were calculated for an isotropic and an anisotropic 
clean limit case with $\<a_{\bf k}^2\> = 0.08$ (Fig.~\ref{lanall}).
We note, that the clean limit approximation may, of course, be questioned because 
of the rather high residual resistance repoted in Ref.~\onlinecite{lanpap}
where all data used for the present analysis have been measured on one individual
sample. Supposing the high $\rho_0\simeq 9$\,$\mu\Omega $cm to be intrinsic, we had
to consider an impurity scattering rate $t^+\simeq 10$\,meV which, however, 
did not allow to find any reasonable set of parameters describing the experimental
results on the upper critical field and specific heat.
In the view of the present Eliashberg approach, these data are compatible with
scattering rates $t^+\leq 1$\,meV indicating that the high 
$\rho_0\simeq 9$\,$\mu\Omega $cm of \lan\ comes from scattering on the grain boundaries
rather than from scattering inside the grains which were shown by high resolution
electron microscopy to be practically defect free.\cite{lanpap}

The upper critical magnetic field $H_{c2}(T)$ shown in Fig.~\ref{lanall} 
was fitted with the anisotropy parameter obtained from the analysis of the 
thermodynamic properties, $\<a_{\bf k}^2\> = 0.08$, a mean Fermi 
velocity $\<v_F\> = 0.225\times 10^6$\,m/s and its anisotropy parameter of 
$\<b_{\bf k}^2\> = 0.245$\,. A fit to the experimental data with an anisotropy 
parameter of $\<a_{\bf k}^2\> = 0.06$ is included for comparison. 
The mean Fermi velocity obtained in the analysis lies between the 
root mean square anisotropies 
$\sqrt{\< v_x^2\>} = 0.292\times 10^6$\,m/s and $\sqrt{\< v_z^2\>} = 0.148\times 10^6$\,m/s 
calculated by Singh and Pickett.\cite{sing} 
In contrast to the borocarbides a consistent description of 
the thermodynamic properties and the upper critical field can also be 
achieved with $s = 0$, which results in an electron-phonon coupling factor 
$\lambda = 0.87$, anisotropy parameters $\< a_{\bf k}^2\> = 0.06$ and 
$\< b_{\bf k}^2\> = 0.27$, and a mean Fermi velocity of $\< v_F\> = 
0.2\times 10^6$\,m/s. 
Different coupling factors $\lambda$ are obtained with different 
coupling functions $\alpha^2(\omega) \sim \omega^s$ ($s=0,-1/2$) because, 
after rescaling the spectral functions to obtain the critical 
temperature of the system and a pseudopotential $\mu^*=0.13$, the 
low-energy contributions to $\alpha^2F(\omega)$ are enhanced with $s = 
-1/2$ compared to the case of $s=0$. 
The coupling factor $\lambda$ obtained with the coupling function 
$\alpha^2(\omega) \sim \omega^0$ is in good agreement with the estimate 
$\lambda \equiv \gamma/\gamma_{bs} -1 = 0.85$. In general, both descriptions 
($ s = 0, -1/2$) require a higher value for the anisotropy parameter 
$\<a_{\bf k}^2\>$ compared to the borocarbides in order to describe the 
experimental data, because the agreement between theory and experiment 
improves with increasing 
$\<a_{\bf k}^2\>$ as is clearly demonstrated by the comparison with the 
specific-heat difference, the thermodynamic critical field, and the 
deviation function of \lan\ shown in Fig.~\ref{lanall}. The rather 
large anisotropy of $\< a_{\bf k}^2\> \simeq 0.08$ further explains the 
temperature dependence of 
the upper critical field and $H_{c2}(0)$ of $\sim 8$\,T comparable to that 
of \luni, since $\< a_{\bf k}^2\>$ has an enhancing effect on $H_{c2}$.
  
\subsection{Comparison of \luy\ and \lan}

The most distinctive feature of the borocarbide and boronitride
superconductors pointed out in previous 
papers\cite{lanpap}$^,$\cite{hils} is the temperature dependence
of the electronic specific heat in the superconducting state: While Y- 
and \luni\ exhibit an almost cubic temperature dependence of $C_{eS}(T)$, 
\lan\ shows an approximately exponential behavior of $C_{eS}(T)$ close to 
the BCS predictions. Thus, \lan\ was supposed to be a weak-coupling BCS 
superconductor, because other characteristics of the 
superconducting state of the boronitride like the thermodynamic ratios 
(e.g. $\Delta C/\gamma T_c=1.4$) and the deviation function $D(t)$ (see 
Fig.~\ref{devivgl}) were 
also found to be much closer to the BCS predictions than those of the 
borocarbides. 
The above analysis of the thermodynamic properties and the 
upper critical field, however, demonstrates that the difference between 
the thermodynamic ratios and the deviation function of the borocarbides and 
the boronitride arises from a larger 
anisotropy of the electron-phonon interaction in the latter rather than 
from a smaller coupling strength ($\lambda\simeq 0.87$ or $1.02$ 
for \lan\ compared to $\lambda \simeq 1.02-1.22$ for \luy). The different 
temperature dependence of the heat capacity in the superconducting state -
cubic in case of the borocarbides - cannot be explained by our Eliashberg 
calculations
which fit the specific heat difference $\Delta C = C_s-C_n$, but not the 
absolute $C_{eS}(T)$, of
e.g. \yni\ depicted in Fig.~\ref{cp_log} as a semilogarithmic plot of 
$C_{eS}/\gamma T_c$ versus the 
inverse reduced temperature $T_c/T$. As the gap obtained by the above 
calculations is free of nodes, it yields an exponential 
temperature dependence for $C_{eS}(T)$ which becomes linear in the 
semilogarithmic representation. This is in contrast to the experimental 
result 
for \yni\ which exhibits a pronounced curvature in this plot indicative 
for a power law. In section 
IV C we noted, that a satisfactory description of the thermodynamic 
properties of the borocarbides and in particular their thermodynamic 
critical 
field is only achieved when the calculations are performed with gamma values 
of about 5\% smaller than the experimental values. 
By adding a linear contribution corresponding to the 5\% 
of the normal state gamma value ($1.1$\,mJ/mol\,K$^2$) omitted before, 
an improvement of the temperature dependence of the calculated 
$C_{eS}(T)$ is achieved (dashed line in Fig.~\ref{cp_log}). These additional 
electronic
contributions may be attributed to a normal state impurity phase and/or to
contributions from the small electron pocket, for which Terashima 
{\it et al.\/}\cite{terashima} proposed a significantly reduced gap. The 
latter argument is supported by the specific heat results obtained on a 
single crystal \yni\ by Nohara {\it et al.}\cite{nohara}  which are included 
in Fig.~\ref{cp_log} for comparison. For
temperatures $T_c/T$ between 1 and 5 (i.e. $T=15.5$-$3$\,K) both 
experimental 
data sets are in rather good agreement, but clearly deviating from our
calculations. Below about 3\,K an exponential temperature dependence can 
be resolved from the data of Nohara {\it et al.\/}\cite{nohara}. This may be 
attributed to the 
small gap on the electron pocket which can be identified as the third Fermi 
surface sheet 
detected by Dugdale {\it et al.\/}\cite{dugdale}$^,$\cite{terashima} The 
third Fermi surface sheet was not 
considered in our calculations because its contribution appeared to be 
negligible in $C_s-C_n$. The exponential low temperature 
behavior of $C_{eS}$ in particular for the polycrystalline \yni\ is covered 
by a linear electronic contribution from small amounts of a secondary 
phase of about $0.4(2)$\,mJ/mol\,K$^2$. Although this contribution is 
significantly reduced in the samples of the pseudoquaternary system, the 
overall feature of the semilogarithmic plot is preserved. Accordingly the 
conclusions concerning the reduced gap on the third Fermi surface sheet 
can also be drawn for \luy.

\section{ CONCLUSIONS}

Model phonon spectra obtained from specific heat measurements, rescaled to 
give the critical temperatures of the system and for a 
Coulomb-pseudopotential value 
of $\mu^* = 0.13$, resulted in coupling strengths $\lambda$ ranging from 
$1.02$ for \yni\ to $1.22$ for \luni. We were able to show that a 
consistent description of the thermodynamic properties and the upper 
critical 
field of the series \luy\ and the boronitride \lan\ within the Eliashberg 
theory is achieved only, if anisotropy effects of the electron-phonon 
coupling and of the Fermi velocity are included. Excellent agreement 
between theory and 
experiment was achieved in the analysis of the series \luy\ with
anisotropy parameters $\<a_{\bf k}^2\>=0.02$--$0.03$ and 
$\<b_{\bf k}^2\>=0.245$--$0.3$.  
The model spectral functions for \lan\ yielded coupling factors $\lambda = 
1.02$ ($s = -1/2$) and $\lambda = 0.87$ ($s = 0$) comparable to those of 
\luy. The thermodynamic properties and 
the upper critical field are well 
described with $\<a_{\bf k}^2\> = 0.08$ and $\<b_{\bf k}^2\> = 
0.245$ ($s = -1/2$) or $\<a_{\bf k}^2\> = 0.06$ and $\<b_{\bf k}^2\> = 
0.27$ ($s = 0$). For \lan\ it is of importance to include anisotropy in 
order to describe its upper critical field and - in contrast to the 
borocarbides - its thermodynamic properties. 
 
Coupling and impurity effects are not capable of explaining the 
$T_c$-behavior within the series \luy. 
The behavior of $\gamma$ as a function of $x$ suggests $N(0)$, aside from 
$\lambda$, to be the 
reason for the minimum in $T_c(x)$, which is confirmed by an approximate 
calculation of $N(0)$ from $\gamma$. 
The approximately cubic temperature dependence of $C_{eS}(T)$ of 
the borocarbides cannot be 
accounted for by the current Eliashberg calculations due to a small gap 
of the third Fermi surface sheet. This contribution, however, is 
suppressed in $\Delta C$ and therefore the Eliashberg calculations are in 
good agreement with the related experimental thermodynamic properties. 

\acknowledgments
We thank M. Nohara for providing the specific heat data included in 
Fig.~\ref{cp_log}. 
This work was supported by the Austrian Science Foundation under Grant 
No.~P~11090 and by the K\"arntner Elektrizit\"atsgesellschaft (KELAG).

%%%%%%%%%%%%%%%%%%%%%% Table 1
\begin{table}                    
\caption{Parameters of the model spectrum}
\begin{center}
\begin{tabular}{l|c|c|c|c|c} 
 cont. & 
 $\Omega_{D} $ &
 $\Omega_{E_0} $ &
 $\Omega_{E_1} $ &
 $\Omega_{E_2} $ &
 $\Omega_{E_3} $ (cut-off) \\ \hline
 & free & free & free & free & 103.4\,meV (fixed) \\ 
 $\sigma_i$  (meV) & { } -- { } & 1.6 & 1.6 & 12.1 & 25.9 \\
 $g_i$ & 3 & 0.3 & 1.2 & 8.5 & 5 
\end{tabular}
\end{center}
\label{herwtab}
\end{table}
%%%%%%%%%%%%%%%%%%%%%% Table 2 %%%%%%%%%%%%%%%%%%%%%%%%%%%%%%%
\begin{table}
\caption{Parameters used for the calculation of thermodynamic
properties of \luy. The mass enhancement factor or
electron-phonon coupling factor, $\lambda$, the number of atoms in
a certain volume, $n_A$, and the Sommerfeld constant used in the 
theoretical description, $\gamma_c$, are compared within the series \luy.} 
\begin{center}
\begin{tabular}{c|c|c|c|c}
 $x$ & $T_c$ [K] & $\lambda$ & $n_A$ [$10^{22}/{\rm cm}^3$] & $\gamma_c$
[mJ/mol 
K$^2$] \\\hline
\hline
0 & 16.01 & 1.220 & 9.405 & 19.7\\
0.1 & 15.56 & 1.182 & 9.382 & 18.9\\
0.5 & 14.90 & 1.058 & 9.263 & 18.1 \\
0.8 & 14.68 & 1.018 & 9.220 & 18.0 \\
1 & 15.45 & 1.017 & 9.175 & 18.6 \\
\end{tabular}
\end{center}
\label{table2}
\end{table}

%%%%%%%%%%%%%%%%%% Fig 1
\begin{figure}
\caption{Influence of different weights for the two Fermi-surface 
sheets on the upper critical field $H_{c2}(T)$. The numerical result in 
the case of equal weights (1:1) is a fit to the upper critical field of 
\luni\ with $\< a^2 \>=0.02,\ \< b^2\> = 0.25,\ {\rm and}\ \< v_F\> = 
0.280\times 10^6$\,m/s. $D_{H_{c2}}(t)$ is depicted in the small insert.}
        \label{luisoan}
\end{figure}
%%%%%%%%%%%%%%%%%%%%%%%% Fig 2
\begin{figure}
\caption{Influence of different signs of $a_{\bf k}$ and $b_{\bf 
k}$ on the upper critical field $H_{c2}(T)$. Both Fermi-surface sheets 
were weighed with the same factor (1:1). The experimental result for 
\luni\ is labeled with open circles.}
        \label{luiso}
\end{figure}
%%%%%%%%%%%%%%%%%%%%%%% Fig 3
\begin{figure}
\caption{$C_p/T$ {\it vs.\/} $T$ plot for \luy\/ at zero 
external field.}
        \label{ylunull}
\end{figure}
%%%%%%%%%%%%%%%%%% Fig 4 and 5
\begin{figure}
\caption{$C_p/T$ {\it vs.\/} $T^2$ plot for \luy\/ at an external 
field of $B=9$\,T. The straight dashed lines indicate the extrapolation of 
the data to $T\rightarrow 0$.}
        \label{ylucel}
\end{figure}
\begin{figure}
\caption{Normal-state contribution to the specific heat of \luy.}
        \label{yluphon}
\end{figure}
%%%%%%%%%%%%%%%%%%%%%%%% Fig 6 %%%%%%%%%%%%%%%%%%%
\begin{figure}
\caption{Model phonon spectra obtained from the phonon 
contribution to the specific heat. The spectral densities 
$\alpha^2F(\omega)$ used in the analysis of \luy\ were obtained by 
applying a coupling function $\alpha^2(\omega)\sim \omega^{-1/2}$ to the 
model PDOS. The three dimensional plot shows the evolution of modes {\it 
vs.} $x$ of $F(\omega)$.}
        \label{spek}
\end{figure}
%%%%%%%%%%%%%%%%%%%%%% Fig 7 %%%%%%%%%%%%%%%%%
\begin{figure}
\caption{Sommerfeld constants used in the theoretical 
description compared to those obtained 
from specific heat measurements. }
        \label{gamma}
\end{figure}
%%%%%%%%%%%%%%%%%%%%% Fig 8 %%%%%%%%%%%%%%%%%%%%%%
\begin{figure}
\caption{Specific heat difference $\Delta C(T) = C_s(T)-C_n(T)$ (a),
thermodynamic critical field $H_c(T)$ (b), and deviation function
$D(T)$ (c) of \luni\ compared to theoretical
calculations for an isotropic and an anisotropic case with $\<a_{\bf k}^2\> 
= 0.03$. The upper critical field (d) is a fit to the experiment with 
anisotropy parameters $\<a_{\bf k}^2\> = 0.02$ and $0.03$.} \label{luall} 
\end{figure} 
%%%%%%%%%%%%%%%%%% Fig 9
\begin{figure}
\caption{Specific heat difference $\Delta C(T) = C_s(T)-C_n(T)$,
of \luy\ with $x=0.1,\ 0.5,\ 0.8$, and $1$ compared to theoretical 
calculations for an isotropic and an anisotropic case with $\<a_{\bf 
k}^2\> = 0.03$.} \label{cps} 
\end{figure}
%%%%%%%%%%%%%% Fig 10 and 11 and 12
\begin{figure}
\caption{Thermodynamic critical field $H_c(T)$
for the samples with $x=0.1,\ 0.5,\ 0.8$, and $1$ compared to theoretical
calculations for an isotropic and an anisotropic case with $\<a_{\bf k}^2\> 
= 0.03$.} 
\label{hcs} \end{figure}
\begin{figure}
\caption{Upper critical field $H_{c2}$ of \luy\ with $x=0.1,\ 0.5,\ 0.8$, 
and $1$ compared to theoretical 
calculations for anisotropic cases with $\<a_{\bf k}^2\> 
= 0.02$ and $0.03$.} 
\label{hc2s} \end{figure}
\begin{figure}
\caption{Deviation function
$D(T/T_c)$ of \luy\ with $x=0.1,\ 0.5,\ 0.8$, and $1$ compared to theoretical
calculations for an isotropic and an anisotropic case with $\<a_{\bf k}^2\> 
= 0.03$.} 
\label{devis} \end{figure}
%%%%%%%%%%%%%%%%%%%%%%%%%%%%%%%%%%%%%%%%% Fig 13
\begin{figure}
\caption{The critical temperature $T_c$, electron-phonon coupling 
strength $\lambda$, Sommerfeld constant $\gamma$, and density of states 
$N(0)$ as a function of $x$ for the series \luy.} 
\label{tcargh} 
\end{figure}
%%%%%%%%%%%%%%%%%%%%%%%%%%% Fig 14 %%%%%%%%%%%%%%%%%%%%
\begin{figure}
\caption{Specific heat difference $\Delta C(T) = C_s(T)-C_n(T)$,
thermodynamic critical field $H_c(T)$, and magnetic deviation function
$D(T)$ of \lan\ compared to theoretical 
calculations for an isotropic and an anisotropic case with $\<a_{\bf k}^2\>
= 0.08$\,. The upper critical field is a fit to the experiment with
anisotropy parameters $\<a_{\bf k}^2\> = 0.06$ and $0.08$\,.} 
\label{lanall} \end{figure}
%%%%%%%%%%%%%%%% Fig 15 %%%%%%%%%%%%%%%%%%%
\begin{figure}
\caption{Spectral density $\alpha^2F(\omega)$ used in the
analysis of \lan\ obtained from a model phonon spectrum, which was rescaled
to give the critical temperature of the sample $T_c=11.73$\,K and for a 
fixed pseudopotential $\mu^* = 0.13$\,.} 
\label{lana2f} \end{figure}
%%%%%%%%%%%%%%%%%%%%%% Fig 16
\begin{figure}
\caption{Deviation functions $D(T/T_c) = D(t)$ {\it vs.\/} $t^2$ of \luy\ 
and \lan. The BCS-result (solid line) was 
added for comparison. } \label{devivgl} \end{figure}
%%%%%%%%%%%%%%%%%%%%%% Fig 17
\begin{figure}
\caption{Electronic specific heat in the superconducting state 
$C_{es}(T)/\gamma T_c$ {\it vs.\/} the inverse reduced temperature 
$T_c/T$ in a semilogarithmic plot. The solid line was calculated by 
adding the normal state contribution $\gamma_c T$ to the 
specific-heat difference between the superconducting and normal state 
$\Delta C$ obtained from the Eliashberg-equations. The data of Nohara 
{\it et al.\/}\cite{nohara} are labeled by filled triangles. } 
\label{cp_log} \end{figure} 
%%%%%%%%%%%%%%%%%%%%%%%%%%%%%%%%%%%%%%%%%%%%%%%%
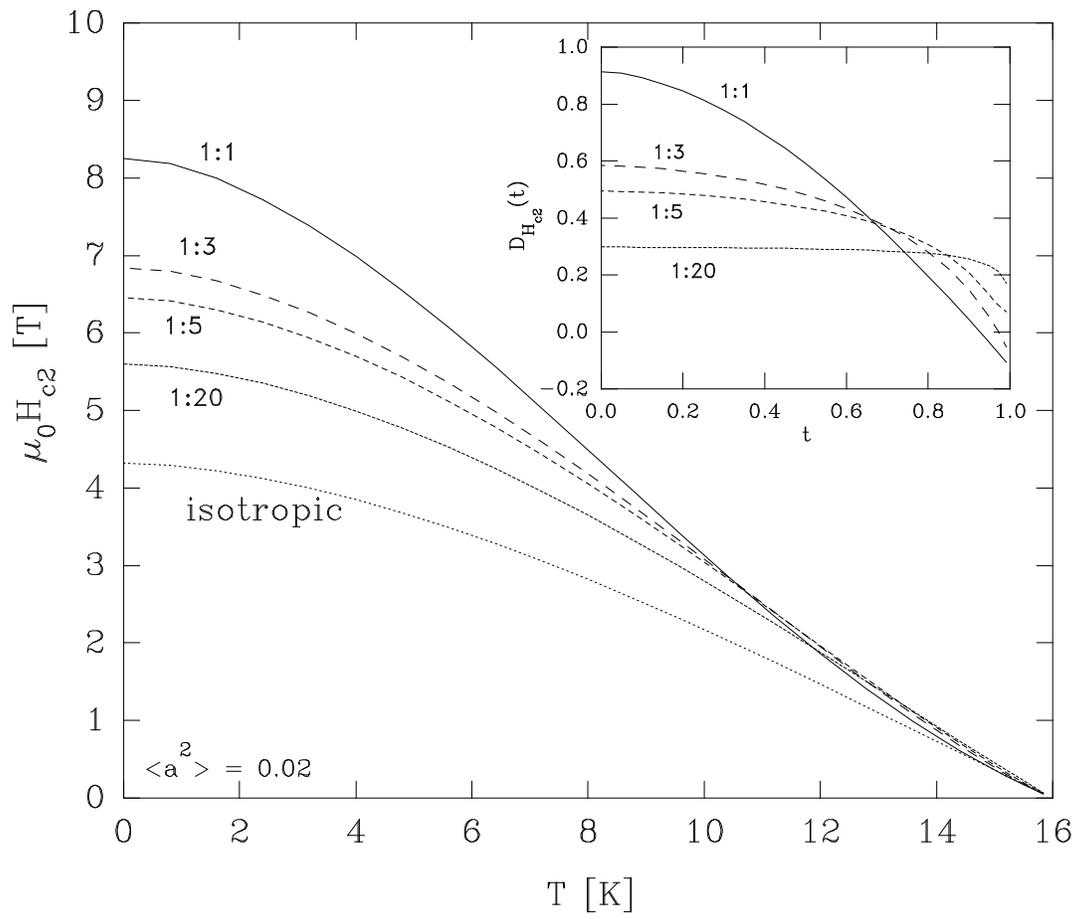
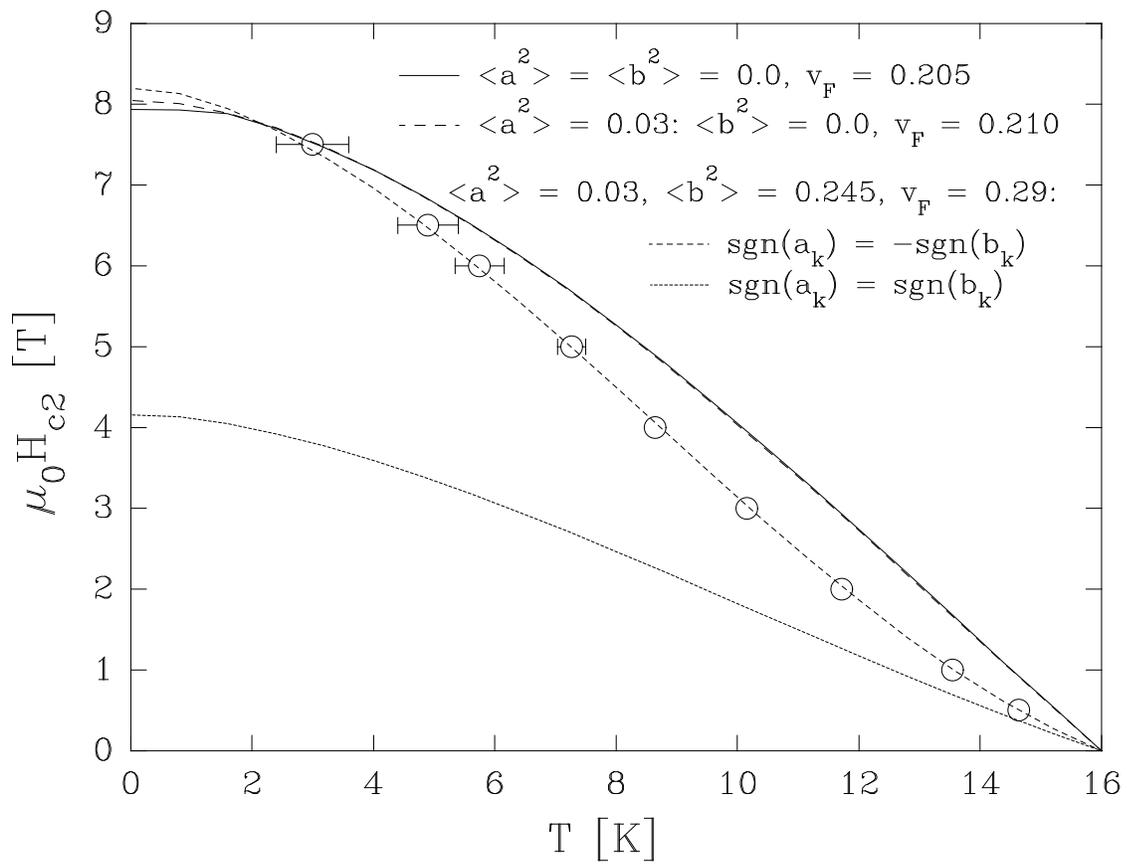
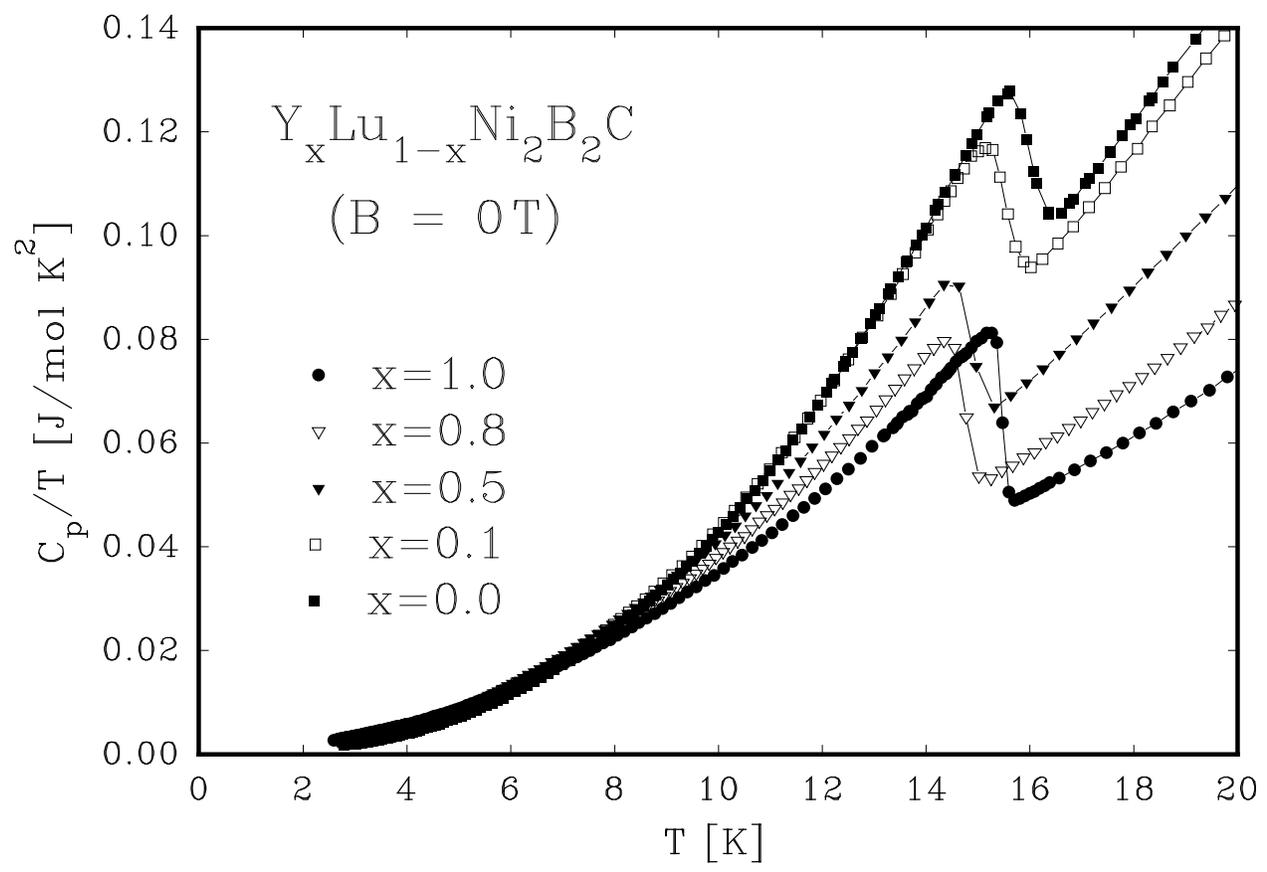
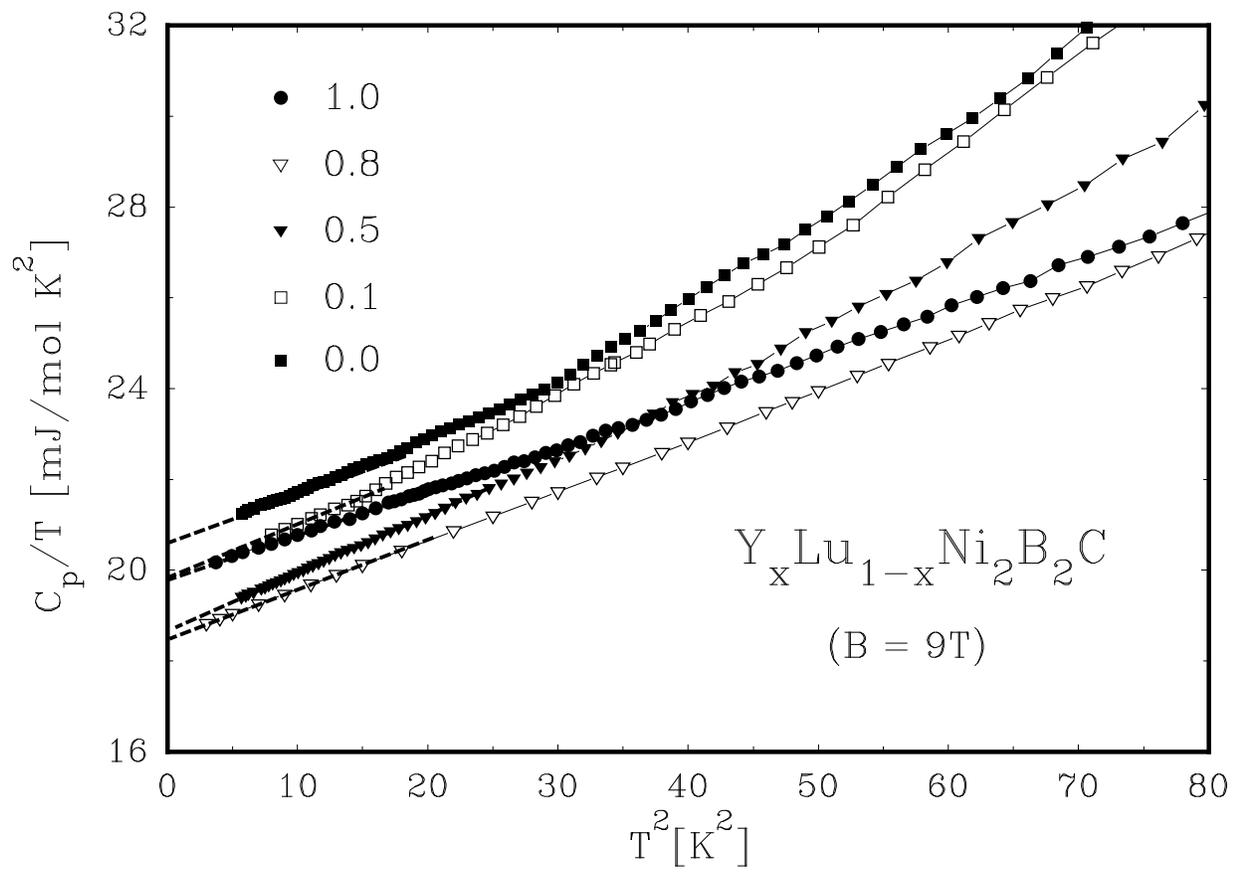
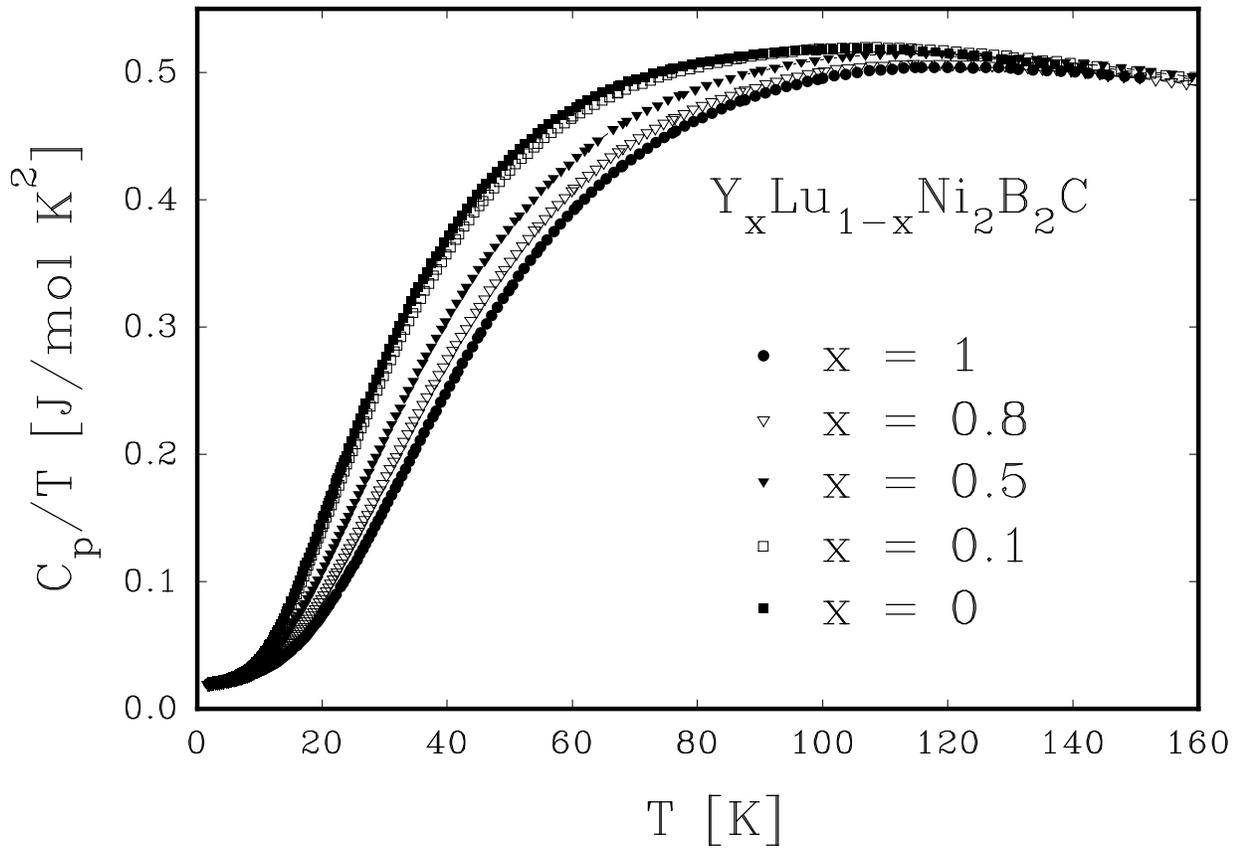
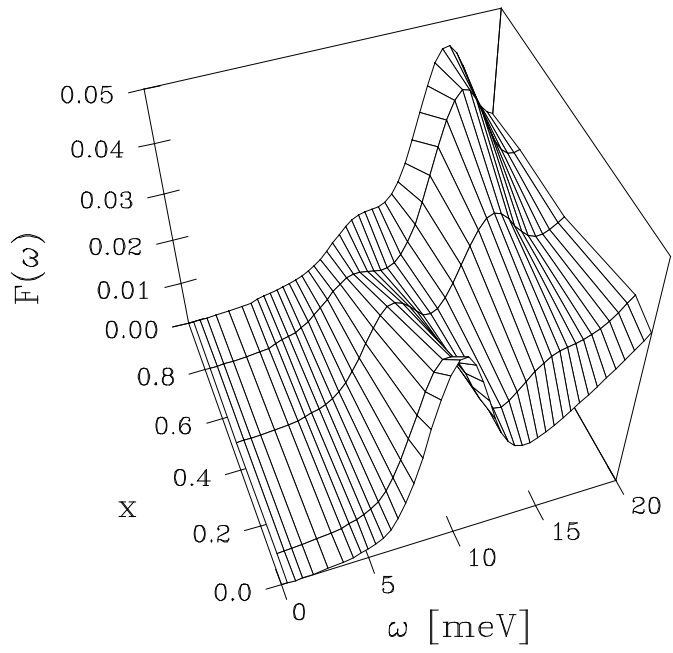
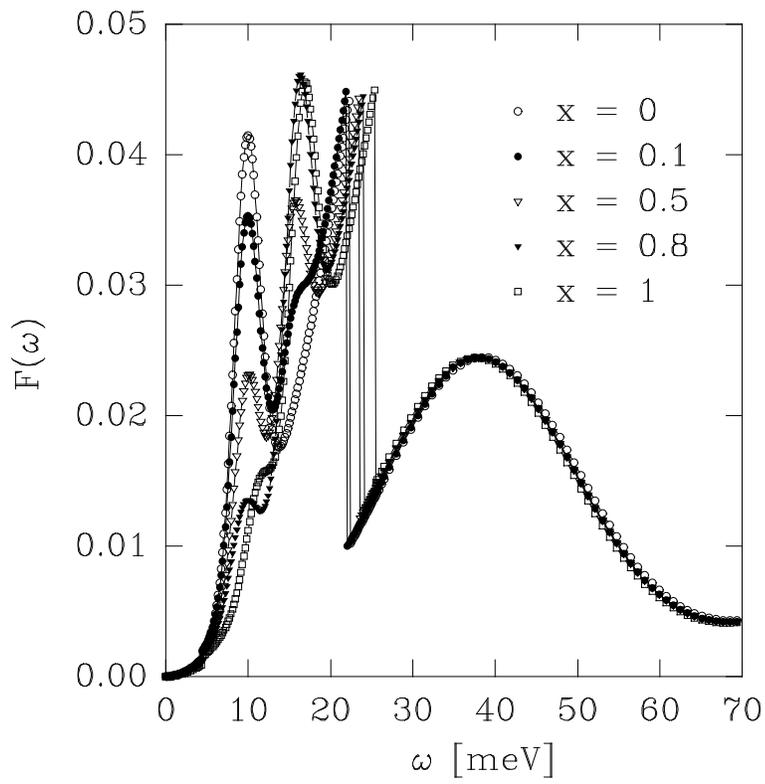
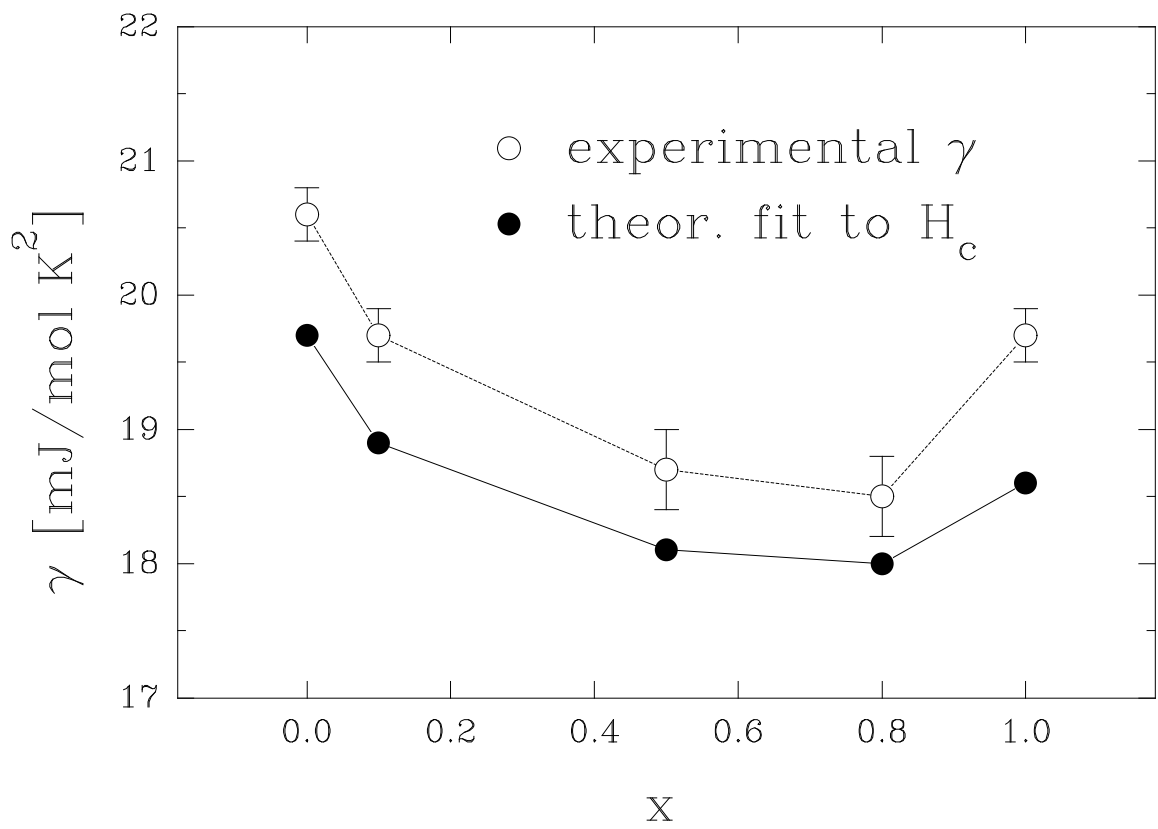
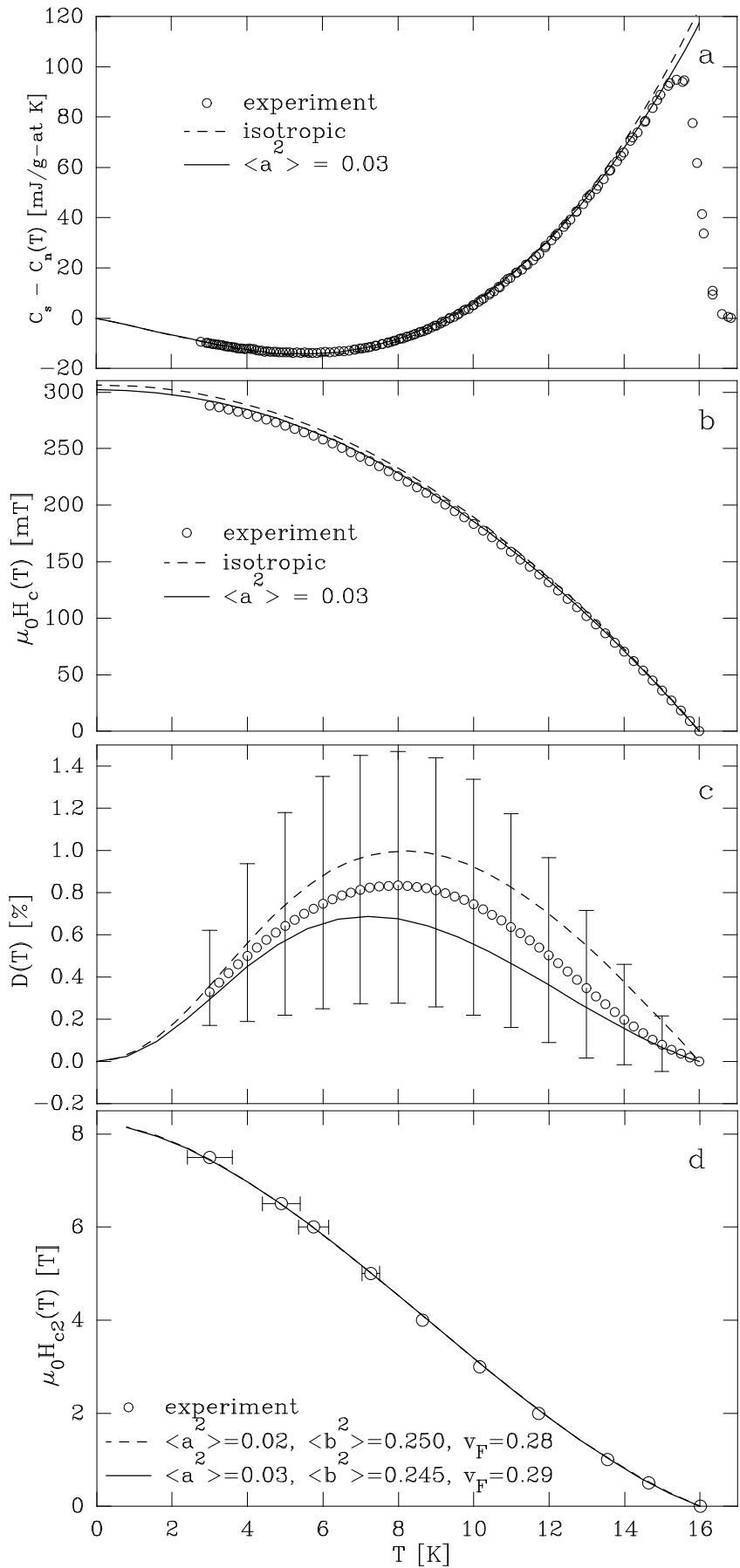
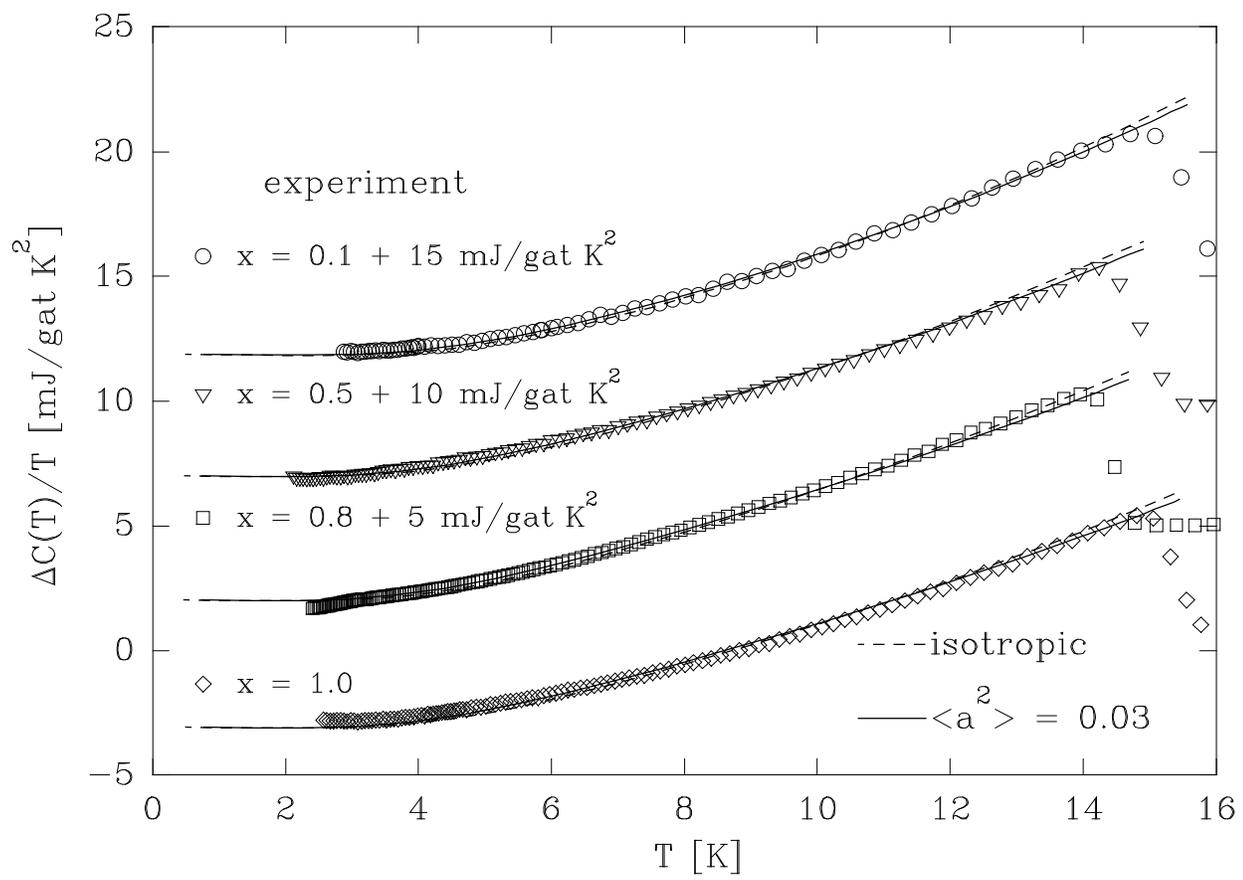
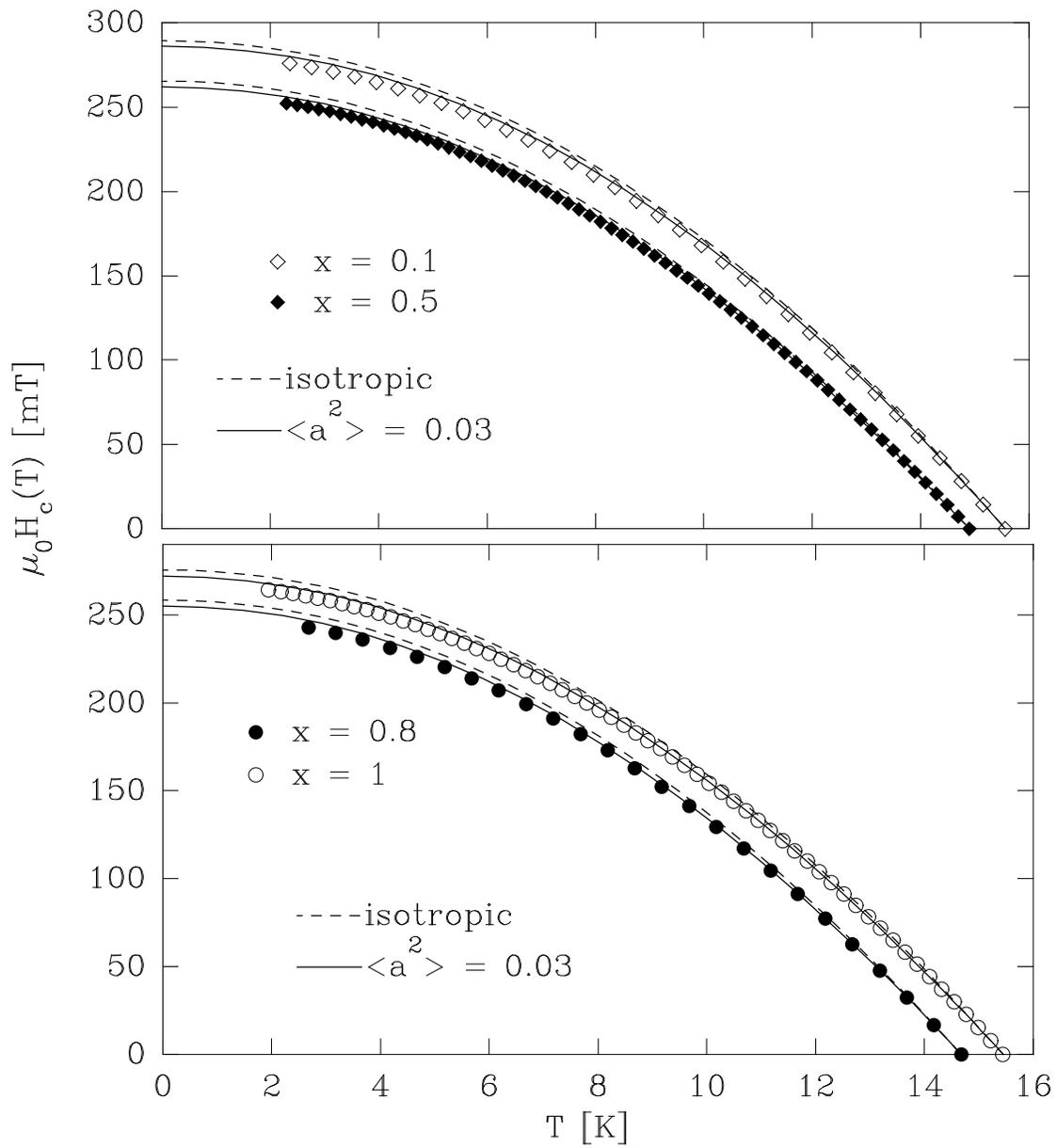
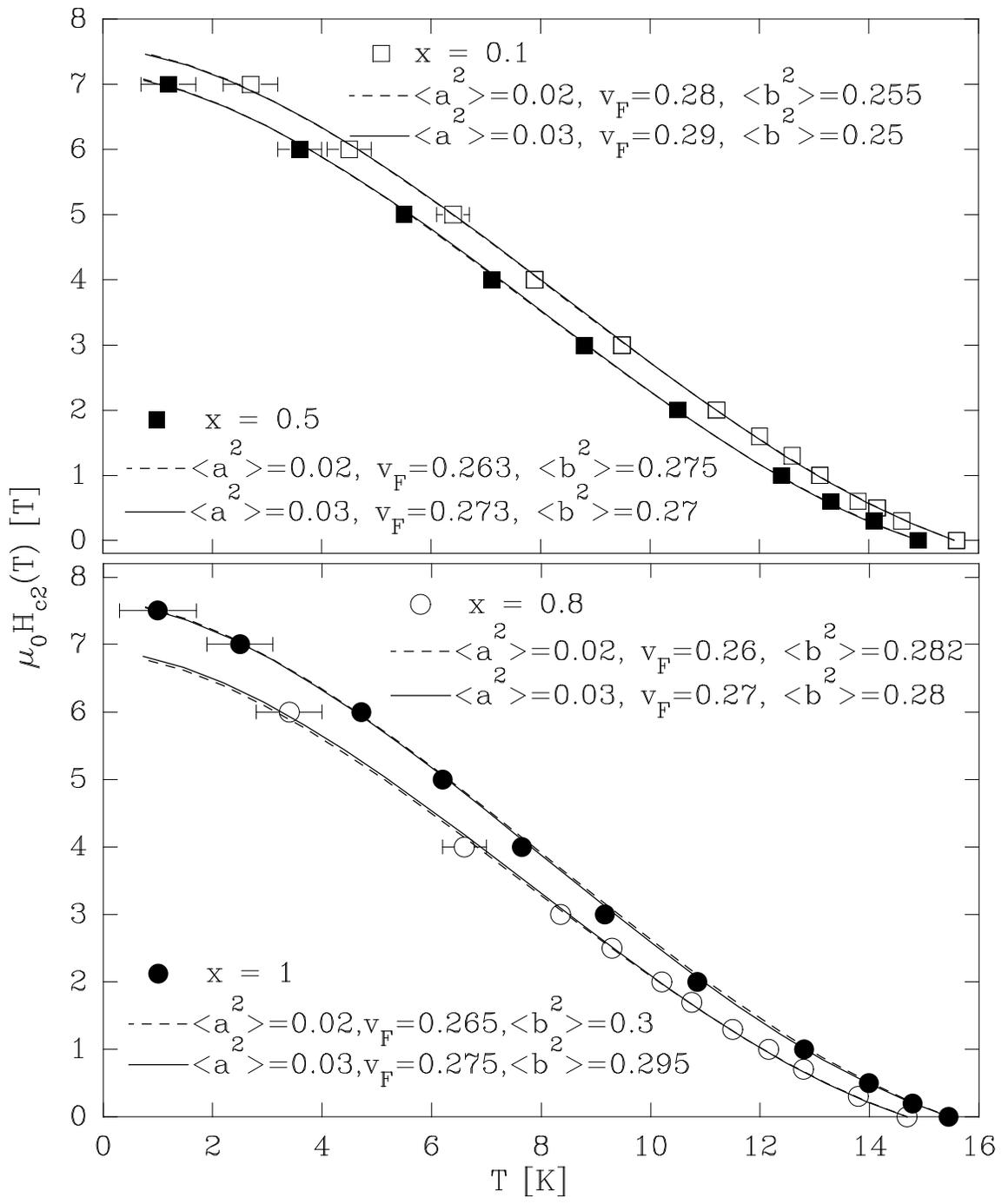
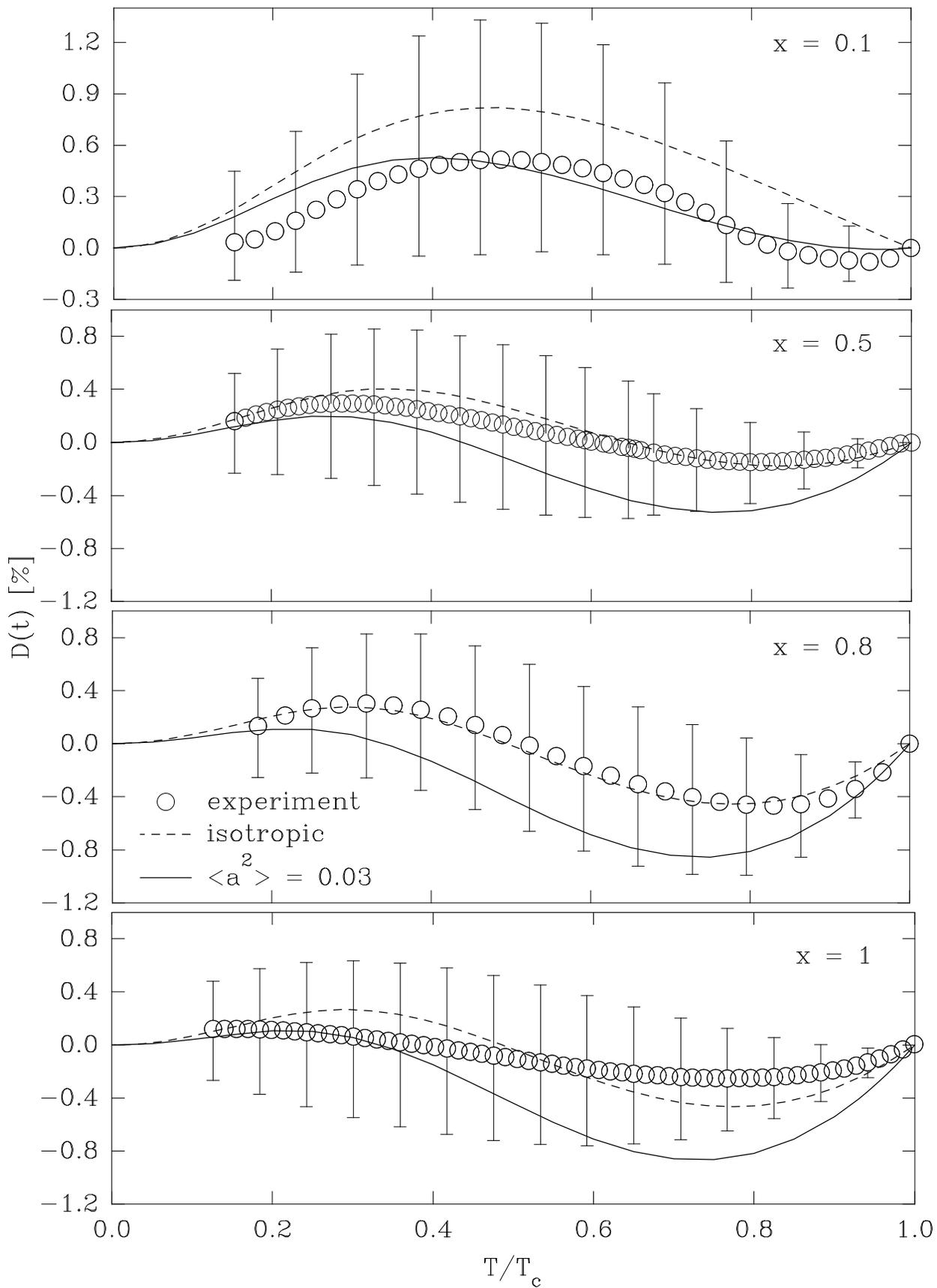
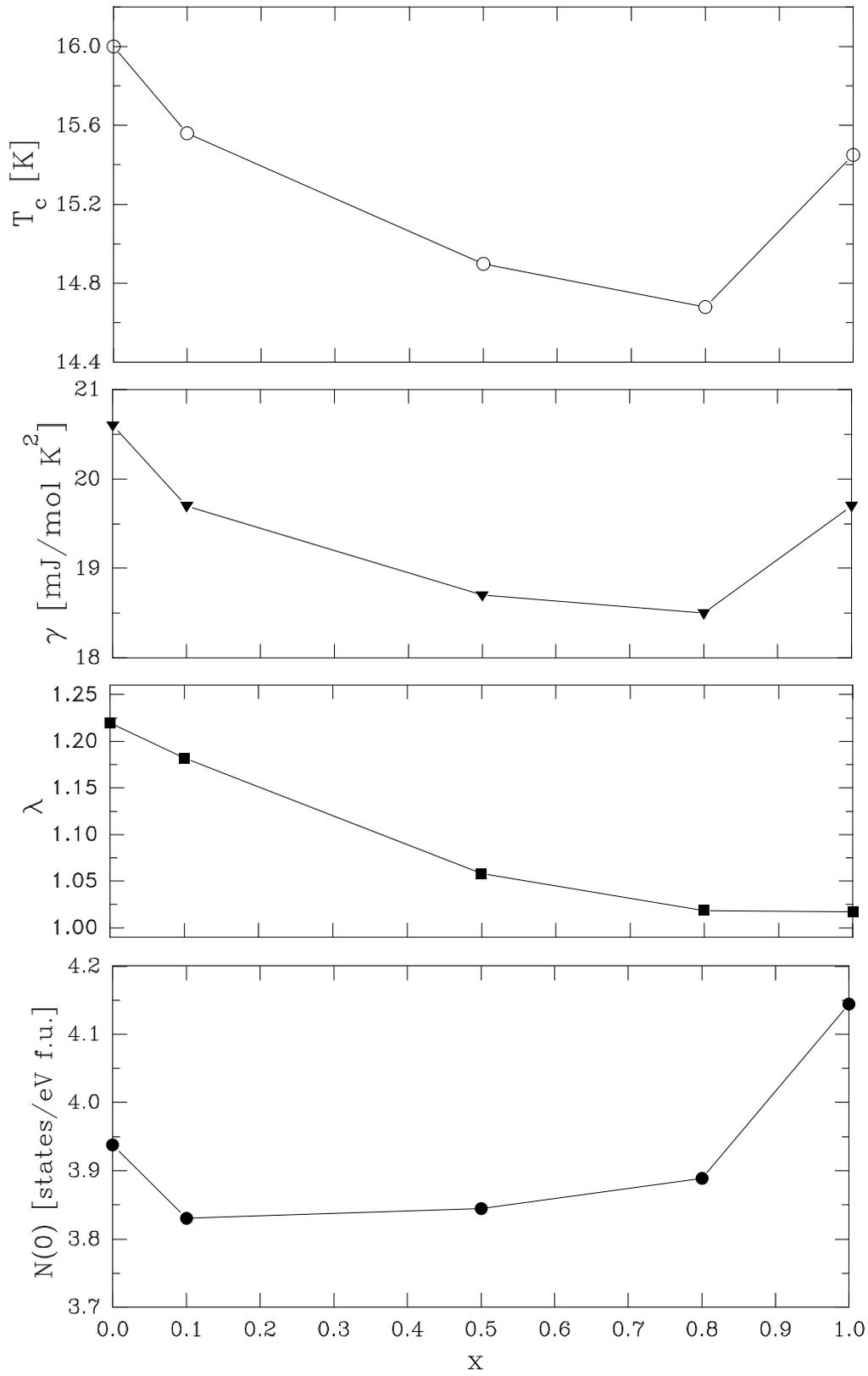
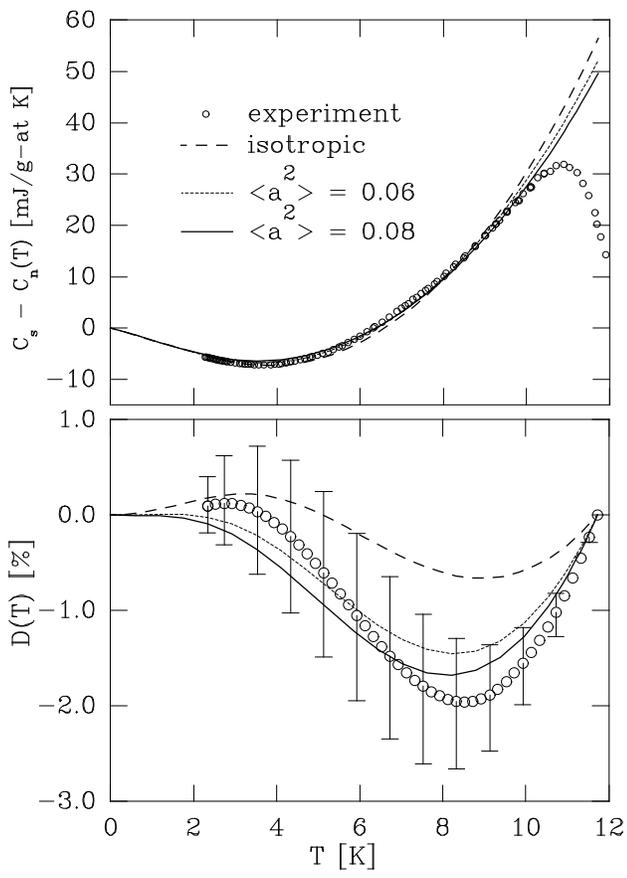
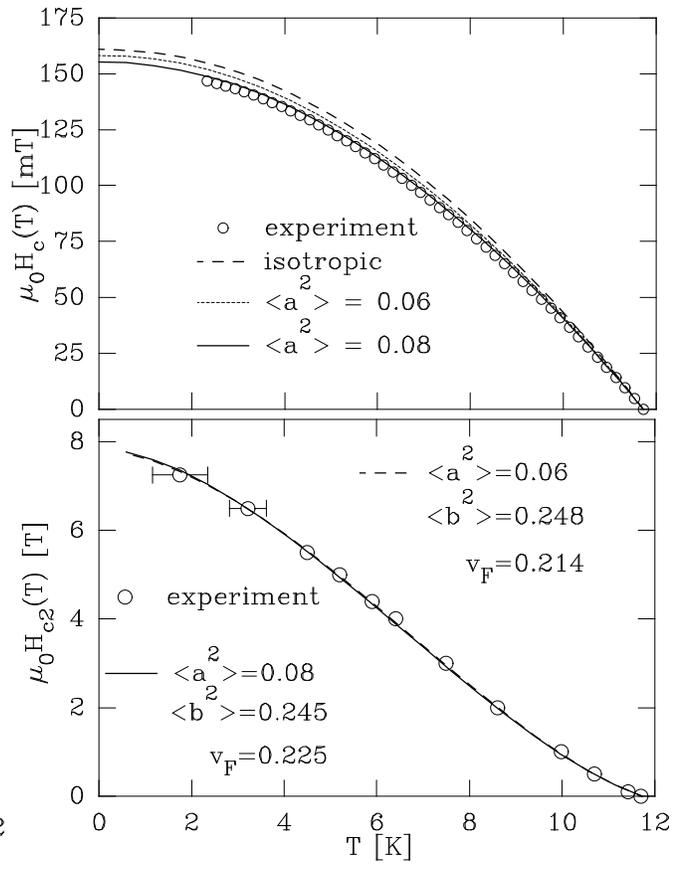
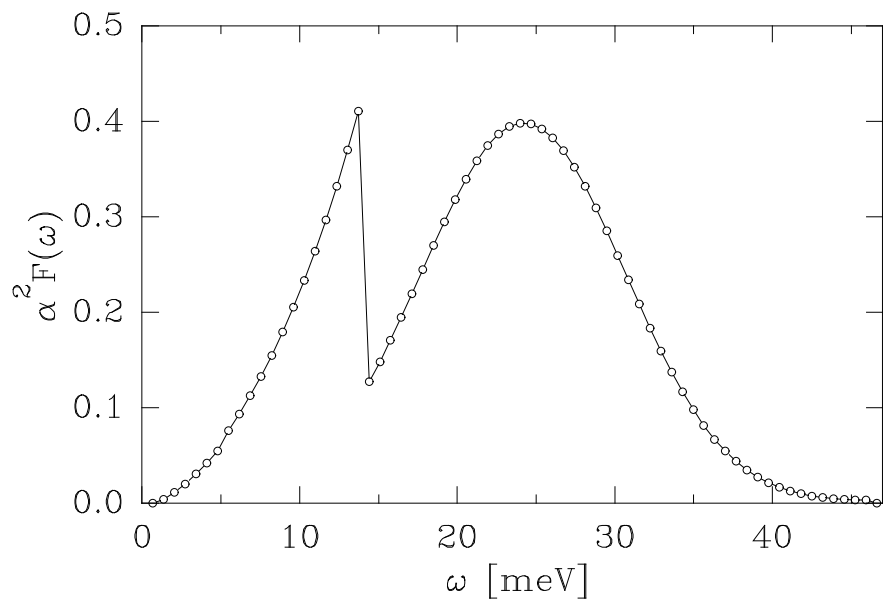
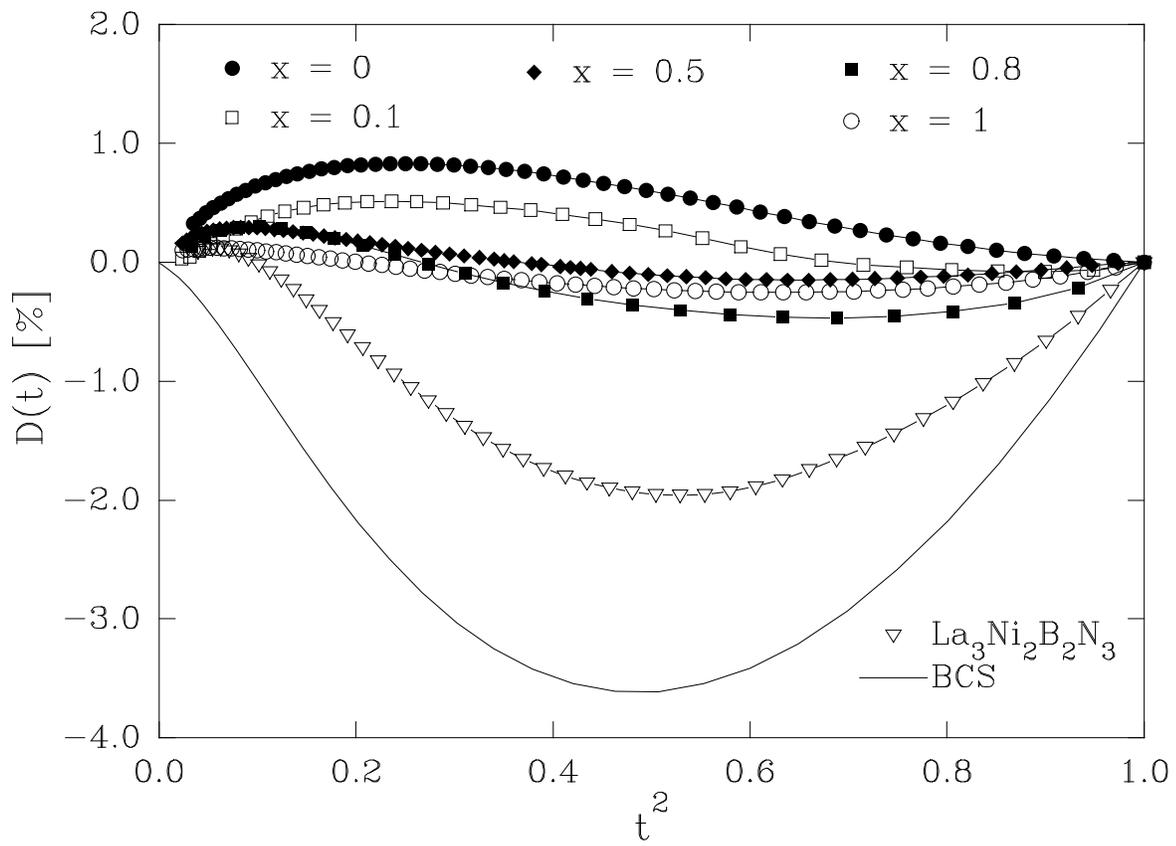
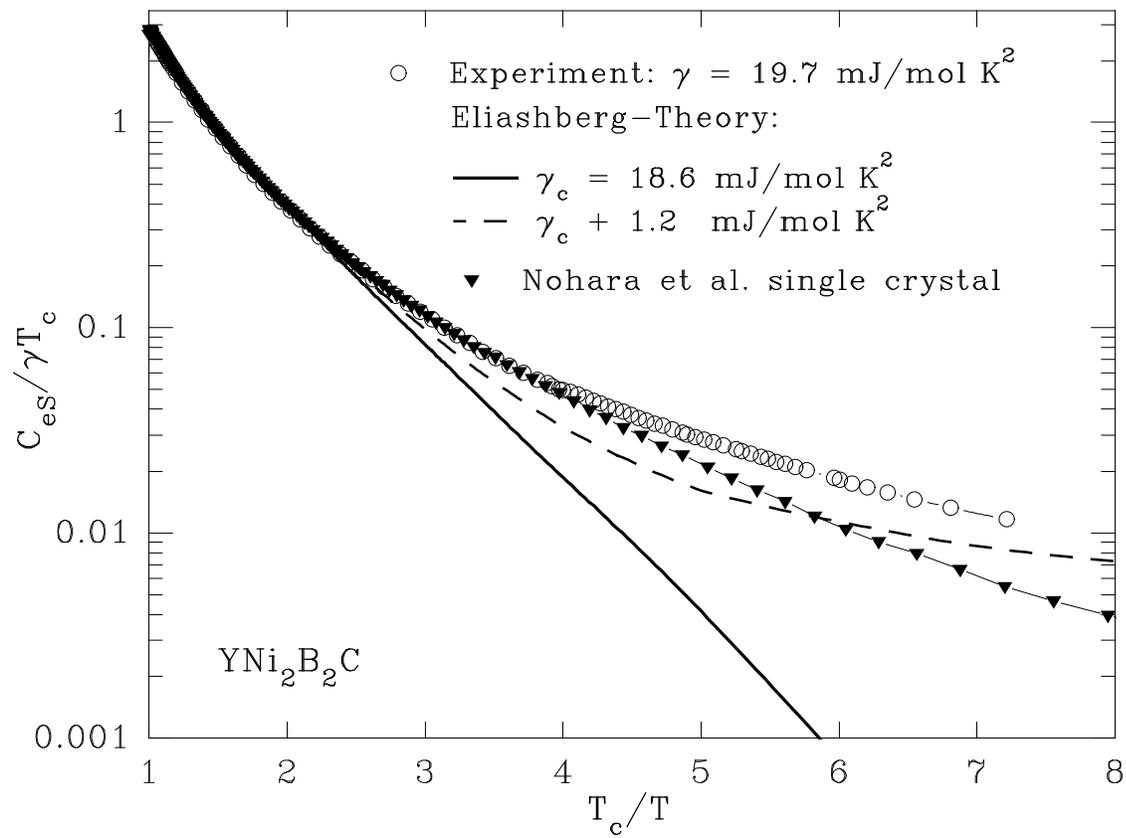
\end{document}